\documentclass[a4paper, 12pt, leqno]{scrartcl}
\pdfoutput=1
\usepackage{%a4wide,
amsmath,amssymb,       %We seem to have used these instead of mathbbol in our Maxwell paper because of arXiv problems
	amsthm,							% theorem enviroment
	bbding,							% ``end of proof'' symbol
bbm,            %We seem to have used this and removed mathbbol from our Maxwell paper because of arXiv problems
	mathrsfs,							% \mathscr	
	MnSymbol,							% mathematics
	slashed,							% Feynman slash
    color
}

\newcommand{\bbone}{{\mathbbm{1}}}

\usepackage[left=2.5cm,right=2.5cm,top=2.5cm,bottom=2.5cm]{geometry}
\usepackage[backend=bibtex8, doi=true, eprint=true, style=alphabetic, maxbibnames=99]{biblatex}
\DeclareFieldFormat[article]{number}{\mkbibparens{#1}}					% sets number in parentheses
\DeclareFieldFormat{pages}{#1}									% removes pp.
\renewbibmacro*{in:}{}										% removes "in:"
\renewbibmacro*{volume+number+eid}{							% bold volume and removes the dot between volume and number
	\textbf{\printfield{volume}}
	\setunit*{\addnbthinspace}
	\printfield{number}
	\setunit{\addcomma\space}
	\printfield{eid}}
\bibliography{bib}
\newtheorem{prop}{Proposition}[section]
\newtheorem{thm}[prop]{Theorem}
\newtheorem{lem}[prop]{Lemma}

\DeclareMathOperator{\id}{id}

\DeclareMathOperator{\pr}{pr}

\DeclareMathOperator{\vol}{vol}
\DeclareMathOperator{\iu}{i}

\renewcommand{\sp}{\mathrm{sp}}
\newcommand{\cosp}{\mathrm{cosp}}

\newcommand{\ip}[2]{{\langle #1\mid #2\rangle}}
\begin{document}
%=====================================================================================================================================================================

%=====================================================================================================================================================================
%title page
%=====================================================================================================================================================================
\title{Pure quasifree states of the Dirac field from the fermionic projector}
\author{Christopher J. Fewster\footnote{chris.fewster@york.ac.uk}\enspace and Benjamin Lang\footnote{bl620@york.ac.uk}\\
 Department of Mathematics, University of York,\\
 Heslington, York YO10 5DD, U.K.}
\date{\today}
\maketitle
%=====================================================================================================================================================================

%=====================================================================================================================================================================
%abstract
%=====================================================================================================================================================================
\begin{abstract} 
We consider the quantised free Dirac field on 
oriented and globally hyperbolic ultrastatic slab spacetimes with compact spatial section and demonstrate how a gauge invariant, pure and quasifree state on the $C$*-completion of the self-dual CAR-algebra can be extracted from the fermionic projector construction of Finster and Reintjes. This state is analogous
to the `SJ-state' of the free scalar field recently discussed in the literature. We prove that this state generically fails to be Hadamard. However, we also show how
a modified version of the construction, inspired by work of Brum and Fredenhagen, yields states which are Hadamard.  
We also relate the Hadamard condition to the finiteness of 
fluctuations of Wick polynomials.
\end{abstract}
%=====================================================================================================================================================================

%=====================================================================================================================================================================
\section{Introduction}

Quantum field theory (QFT) in Minkowski space, in most formulations, is tightly structured around the existence of a Poincar\'e invariant vacuum state.  However, attempts to define similarly natural states in curved spacetime QFT have met with failure, and indeed there is a no-go theorem to the effect that no natural choice is possible under suitable conditions~\cite[\S 6.3]{SPASs12}. 

One of the main conditions of the
no-go theorem is that the putative natural state should depend in a local
fashion upon the spacetime geometry. This leaves open the possibility that
there might be interesting states determined \emph{nonlocally} by the geometry. 
Such a proposal was made recently by Afshordi, Aslanbeigi and Sorkin
for the real scalar field~\cite{AAS12}, under the name `SJ-state'.  
In brief, the idea is to use the advanced-minus-retarded fundamental
solution to determine a self-adjoint operator on the Hilbert space
of square-integrable functions on spacetime, with respect to the 
standard volume measure. The positive part of this operator is then
used to determine the two-point function of a state. One may
give precise conditions under which this prescription does indeed
yield a pure quasifree state~\cite{FeV12b}, which is certainly
independent of any choices of coordinates. However, by explicit
computation, it is known that the SJ-state has a number of unphysical
aspects~\cite{FeV12b,FeV13}: in particular, the SJ-state on
a generic ultrastatic slab with compact Cauchy surface fails to be Hadamard. Here, an ultrastatic slab is a spacetime $(a,b)\times\Sigma$
with a product metric $g=dt\otimes dt -\pr_2^*h$, where
$h$ is a fixed Riemannian metric on $\Sigma$ and $\pr_2$ is
the projection $\pr_2(t,\sigma)=\sigma$. When we say that
the Hadamard condition fails generically we mean, more precisely,
the following: To each $(\Sigma,h)$
with $\Sigma$ compact, one may associate a subset $\mathcal{E}(\Sigma,h)\subset (0,\infty)$ which has Lebesgue measure zero 
and is known to be empty for at least some choices of $(\Sigma,h)$);
the SJ state on $((a,b)\times\Sigma,g)$ is known not to be Hadamard 
provided $b-a\notin \mathcal{E}(\Sigma,h)$. (The result of~\cite{FeV12b} leaves open the issue of whether or not
the SJ state is Hadamard if $b-a\in \mathcal{E}(\Sigma,h)$.)
 
Brum and Fredenhagen showed, however, that the 
SJ prescription may be modified so as to yield Hadamard states which we call \emph{BF-states}~\cite{BF14}. 
This comes at a price: the BF-states are specified not only by the spacetime geometry, 
but also by a smooth compactly supported function on spacetime.   Nonetheless, this provides an interesting and novel construction
of a class of physically acceptable states for the real scalar field. 

In this paper, we consider similar questions for the free
Dirac field. As we describe, there is a direct analogue of the SJ 
construction which is closely related to the `fermionic projector'
programme of Finster (an idea known to Finster and also noted
e.g., in~\cite{BF14,FeV13}; however, we give the first worked out implementation). 
We will call the resulting states \emph{unsoftened FP-states} (for fermionic projector). Following the description of fermionic projectors
set out in~\cite{FR14a}, but adapting it so as to describe states rather than the Dirac sea (the main focus of Finster's programme), we give an explicit computation of the two-point function for the unsoftened FP-state
on an ultrastatic slab, and show that it fails (generically) to be Hadamard.   
However,  
we also describe a modified prescription, inspired by that of \cite{BF14} though somewhat
different in detail, which leads to a class of FP-states, 
parameterised by a choice of nonnegative integrable function. If this
function is smooth and compactly supported, we refer to the
resulting state as a {\em softened FP-state}; the unsoftened FP-state 
results from employing a suitable characteristic function. 
We show that all FP-states are pure, quasifree and gauge-invariant; moreover, all softened FP-states are Hadamard.  We also 
study the fluctuations, in a FP-state, of Wick polynomials defined relative to that state. In the unsoftened
case, the fluctuations are (generically) infinite, again indicating the unphysical nature of unsoftened FP-states. The
fluctuations are finite in the softened FP-states, as is always true of Hadamard states~\cite{BruFreKoe96,DapHacPin09}.
Conversely, we show that any FP-state on an ultrastatic slab spacetime that has finite fluctuations for all its Wick polynomials
is necessarily Hadamard. This is an analogue of a result obtained for the scalar field in~\cite{FeV13}.
%=====================================================================================================================================================================
\enlargethispage{1\baselineskip}
%=====================================================================================================================================================================

%=====================================================================================================================================================================
\section{The quantised free Dirac field on ultrastatic spacetimes and slabs with compact spatial section}\label{sec quantised free Dirac}
%=====================================================================================================================================================================
 
%=====================================================================================================================================================================

%=====================================================================================================================================================================
\subsection{Globally hyperbolic and ultrastatic spacetimes}
%=====================================================================================================================================================================
We define a \emph{spacetime} to be a connected smooth manifold\footnote{For us, a \emph{smooth manifold} is a locally Euclidean and second-countable Hausdorff space with a fixed $\mathcal{C}^\infty$-structure, hence paracompact \cite[Prop.2.24]{Lee03}.} $M$ of dimension $4$ equipped with a Lorentzian metric $g$ of signature $\left(+,-,-,-\right)$ and a time-orientation $\left[T\right]$.  
A \emph{globally hyperbolic spacetime} is a spacetime $\left(M,g,\left[T\right]\right)$ meeting the causality condition and for all $p,q\in M$, $J\left(p,q\right)=J^+\left(p\right)\cap J^-\left(q\right)$, which is the intersection of the causal future of $p$ and the causal past of $q$, is compact. In most of the literature on globally hyperbolic spacetimes, e.g. \cite{Pen72, HE73, Bon83, BEE96}, the strong causality condition is required, however, \cite{BeSa07} has shown that it is enough to merely require the causality condition in the definition of global hyperbolicity.  

A spacetime $\left(M,g,\left[T\right]\right)$ is called \emph{ultrastatic} if it is of smooth product form $M=\mathbb{R}\times\Sigma$ with metric $g=dt\otimes dt-\pr_2^*h$, where $h$ is a Riemannian metric on $\Sigma$ and $\pr_2:\mathbb{R}\times\Sigma\longrightarrow\Sigma$ denotes the projection onto the second factor. Naturally, we will always take the time-orientation such that $\partial/\partial t$ is future-directed. If $M$ is the smooth product manifold $\left(a,b\right)\times\Sigma$, $a,b\in\mathbb{R}$ with $a<b$, and $h$ is a Riemannian metric on $\Sigma$, the spacetime $\left(\left(a,b\right)\times\Sigma,dt\otimes dt-\pr_2^*h,\left[T\right]\right)$ is said to be an \emph{ultrastatic slab}. By \cite[Prop.5.2]{Kay78}, an ultrastatic spacetime or slab is globally hyperbolic if and only if $\left(\Sigma,h\right)$ is a complete Riemannian manifold, as is certainly the case 
by the Hopf-Rinow theorem if $\Sigma$ is taken to be compact \cite[Cor.5.23]{Bon83}. Note that, 
in the terminology of~\cite{FR14a}, our ultrastatic slabs have {\em finite lifetime}. 
%=====================================================================================================================================================================

%=====================================================================================================================================================================
\subsection{The free Dirac equation on globally hyperbolic spacetimes}
%=====================================================================================================================================================================
For a detailed discussion of spin structures, (co)spinors, spin connections and the free Dirac equations, we refer the reader to the literature \cite{Ger68, Ger70, Isham78b, Dim82, FeV02, San08, San10, Fer13b}. Here, we will simply collect the results needed for our purposes.

Let $\textbf{M}=\left(M,g,\left[T\right],\left[\Omega\right]\right)$ be an oriented globally hyperbolic spacetime of dimension $4$, equipped 
with a fixed smooth global Lorentz framing
$\left(\varepsilon_0,\ldots,\varepsilon_3\right)$; that is, the $\varepsilon_\mu$ are smooth vector fields on $M$ such that $\left(\varepsilon_0\left(x\right),\ldots,\varepsilon_3\left(x\right)\right)$ is a time-oriented, oriented and $g$-orthonormal basis of $TM_x$ for each $x\in M$. The dual basis of covector fields will be denoted $\varepsilon^\mu$, 
so that $\varepsilon^\mu(\varepsilon_\nu)=\delta^\mu_\nu$,
and of course $g=\eta_{\mu\nu}\varepsilon^\mu\otimes\varepsilon^\nu$, where
$\eta_{\mu\nu}$ is the standard Minkowski metric in our signature. In this setting, spinor fields may be regarded as $\mathbb{C}^4$-valued smooth functions, i.e., elements in $\mathcal{C}^\infty\left(M,\left(\mathbb{C}^4\right)^*\right)$, while $\left(\mathbb{C}^4\right)^*$-valued smooth functions, i.e., elements in $\mathcal{C}^\infty\left(M,\left(\mathbb{C}^4\right)^*\right)$, are cospinor fields. Elements of $\mathbb{C}^4$ (resp.,  $\left(\mathbb{C}^4\right)^*$) will be regarded as  column (resp., row) vectors. Also note that $\mathcal{C}^\infty\left(M,\mathbb{C}^4\right)$ and $\mathcal{C}^\infty\left(M,\left(\mathbb{C}^4\right)^*\right)$ can be canonically identified with the spaces of smooth cross-sections $\Gamma^\infty\left(\underline{\mathbb{C}^4}_M\right)$ and $\Gamma^\infty\left(\left(\underline{\mathbb{C}^4}_M\right)^*\right)$, where $\underline{\mathbb{C}^4}_M:=\left(M\times\mathbb{C}^4, M, \pr_1, \mathbb{C}^4\right)$ is the trivial smooth complex vector bundle over $M$ of rank $4$ and $\left(\underline{\mathbb{C}^4}_M\right)^*:=\left(M\times\left(\mathbb{C}^4\right)^*, M, \pr_1, \left(\mathbb{C}^4\right)^*\right)$ its dual. The map $\pr_1:M\times\mathbb{C}^4\longrightarrow M$ denotes the projection onto the first factor. 

We choose the Pauli realisation \cite[(7.31)]{BLT75} for the $\gamma$-matrices,
\begin{align*}
&&\gamma^0=\begin{pmatrix}\sigma_0&0\\0&-\sigma_0\end{pmatrix},&&\gamma^i=\begin{pmatrix}0&\sigma_i\\-\sigma_i&0\end{pmatrix}&&\text{and}&&\gamma^5=\gamma^0\gamma^1\gamma^2\gamma^3=-\iu\begin{pmatrix}0&\sigma_0\\\sigma_0&0\end{pmatrix},&&
\end{align*} 
with the Pauli matrices
\begin{align*}
&&\sigma_0=\begin{pmatrix}1&0\\0&1\end{pmatrix},&&\sigma_1=\begin{pmatrix}0&1\\1&0\end{pmatrix}, &&\sigma_2=\begin{pmatrix}0&-i\\i&0\end{pmatrix}&&\text{and}&&\sigma_3=\begin{pmatrix}1&0\\0&-1\end{pmatrix}.&&
\end{align*}
In addition to the Clifford relations $\gamma^\mu\gamma^\nu+\gamma^\nu\gamma^\mu=2\eta^{\mu\nu}$,  
we note the identities $\gamma^0\gamma^\mu\gamma^0=\left(\gamma^\mu\right)^*$, $\left(\gamma^0\right)^*=\gamma^0$ and $\left(\gamma^i\right)^*=-\gamma^i$ (``\,$*$\,'' denotes Hermitean conjugation, i.e., complex conjugation and transposition), which we will use throughout without further mention.

The free Dirac equations for spinors $f\in\mathcal{C}^\infty\left(M,\mathbb{C}^4\right)$ and cospinors $\varphi\in\mathcal{C}^\infty\left(M,\left(\mathbb{C}^4\right)^*\right)$ are now:
\begin{align}\label{Dirac equation spinor}
D^\sp f=\left(-\iu\slashed{\nabla}^\sp+m\right)f=\left(-\iu\gamma^\mu\nabla_{\!\varepsilon_\mu}^\sp+m\right)f=0
\intertext{and}\label{Dirac equation cospinor}
D^\cosp\varphi=\left(\iu\slashed{\nabla}^\cosp+m\right)\varphi=\iu\left(\nabla_{\!\varepsilon_\mu}^\cosp\varphi\right)\gamma^\mu+\varphi m=0,
\end{align}
where  $\nabla^\sp$ and $\nabla^\cosp$ are the \emph{spin connections}, which are given by the expressions
\begin{flalign}\label{spin connections}
&&\nabla^\sp f=df^A\left(\varepsilon_\mu\right)\varepsilon^{\mu}\otimes e_A+\varepsilon^{\mu}\otimes \Gamma_\mu\,f&&\text{and}&&\nabla^\cosp\varphi=
d\varphi_A\left(\varepsilon_\mu\right)\varepsilon^{\mu}\otimes e^{A}-\varepsilon^{\mu}\otimes\varphi\,\Gamma_\mu,&&\\\nonumber
&&&&&&&&\makebox[0pt][r]{$f\in\mathcal{C}^\infty\left(M,\mathbb{C}^4\right)$, $\varphi\in\mathcal{C}^\infty\left(M,\left(\mathbb{C}^4\right)^*\right)$,}
\end{flalign}
where $\Gamma_\mu=\frac{1}{4}\Gamma^\lambda_{\mu\nu}\gamma_\lambda\gamma^\nu$, $\Gamma^\lambda_{\mu\nu}\varepsilon_\lambda=
\nabla\!_{\varepsilon_\mu}\varepsilon_\nu$, $e_A$ is the standard basis for $\mathbb{C}^4$ and
$e^A$ the corresponding dual basis of $(\mathbb{C}^4)^*$. Using Koszul's formula \cite[Thm.3.11]{Bon83}, one can easily show $\Gamma^{\nu}_{\mu\nu}=0$ (no summation!).

As usual, the \emph{Dirac adjoint} is a complex-conjugate linear isomorphism 
\begin{align}\label{Dirac adjoint}
&&\dagger:\mathcal{C}^\infty\left(M,\mathbb{C}^4\right)\longrightarrow\mathcal{C}^\infty\left(M,\left(\mathbb{C}^4\right)^*\right),&&f\longmapsto f^*\gamma^0.&&
\end{align}  
Since it will be clear from the context whether we apply the Dirac adjoint to a spinor or its inverse to a cospinor, we will write $f^\dagger$ and $\varphi^\dagger$ for both the Dirac adjoints of $f\in\mathcal{C}^\infty\left(M,\mathbb{C}^4\right)$ and $\varphi\in\mathcal{C}^\infty\left(M,\left(\mathbb{C}^4\right)^*\right)$. Observe that $\left(\slashed{\nabla}^\sp f\right)^\dagger=\slashed{\nabla}^\cosp f^\dagger$ for all $f\in\mathcal{C}^\infty\left(M,\mathbb{C}^4\right)$ and $\left(\slashed{\nabla}^\cosp\varphi\right)^\dagger=\slashed{\nabla}^\sp\varphi^\dagger$ for all $\varphi\in\mathcal{C}^\infty\left(M,\left(\mathbb{C}^4\right)^*\right)$, which readily yields $\left(D^\sp f\right)^\dagger=D^\cosp f^\dagger$ and $\left(D^\cosp\varphi\right)^\dagger =D^\sp\varphi^\dagger $.

Owing to global hyperbolicity of $\textbf{M}$,  equations 
(\ref{Dirac equation spinor}) and (\ref{Dirac equation cospinor}) have
well-posed Cauchy problems (see \cite[Thm.2.3]{Dim82} or \cite[Thm.2]{Mue11}), and unique retarded and advanced Green's operators
(see \cite[Thm.2.1]{Dim82} or \cite[Thm.1]{Mue11}). We denote the unique retarded and advanced Green operators for spinors (resp., cospinors) by $S^\text{ret},S^\text{adv}$ (resp., $C^\text{ret}, C^\text{adv}$). 
%=====================================================================================================================================================================

%=====================================================================================================================================================================
\subsection{The free Dirac equation on ultrastatic spacetimes and slabs}\label{subsec spatial Dirac operators}
%=====================================================================================================================================================================
Let $\left(\Sigma,h,\left[\Omega\right]\right)$ be an oriented, connected and compact Riemannian manifold of dimension $3$. Hence, $\Sigma$ is parallelisable \cite{Stiefel35/36} and there exist oriented (with respect to $\left[\Omega\right]$) smooth global framings for the tangent bundle and by Gram-Schmidt, the existence of oriented and orthonormal (with respect to $h$) smooth global framings. Fix such a one, say $\left(\eta_1,\eta_2,\eta_3\right)$, then define $\varepsilon_1,\varepsilon_2,\varepsilon_3$ by $\varepsilon_i\left(f\right)\left(t,\cdot\right):=\eta_i\left(f\left(t,\cdot\right)\right)\in\mathcal{C}^\infty\left(\Sigma,\mathbb{R}\right)$ for all $t\in\left(a,b\right)$ and for all $f\in\mathcal{C}^\infty\left(\left(a,b\right)\times\Sigma,\mathbb{R}\right)$, where $-\infty\leq a<b\leq\infty$. 

The quadruple $\textbf{M}=\left(\left(a,b\right)\times\Sigma,dt\otimes dt-\pr_2^*h,\left[\partial/\partial t\right],\left[dt\wedge\pr_2^*\Omega\right]\right)$ is an oriented and globally hyperbolic ultrastatic spacetime (or slab, if $a$ and $b$ are finite) with compact spatial section and the ordered tuple $\left(\varepsilon_0:=\partial/\partial t, \varepsilon_1,\varepsilon_2,\varepsilon_3\right)$ is a smooth global Lorentz framing by construction. We use precisely this smooth global Lorentz framing in our definition of the spin connections and from now on, we will always consider oriented and globally hyperbolic ultrastatic spacetimes or slabs $\textbf{M}$ with spin connections obtained in the way just described. 

Using the Koszul formula \cite[Thm.3.11]{Bon83} one may show that $\Gamma^\lambda_{\mu\nu}$ vanishes if $\mu,\nu$ or $\lambda$ is zero, which implies $\Gamma_0=0$ and $\Gamma_i=\frac{1}{4}\Gamma^k_{ij}\gamma_k\gamma^j$ in (\ref{spin connections}). Furthermore, $\Gamma^k_{ij}$ does not depend on $t\in\left(a,b\right)$ by construction and can be regarded as a smooth function on $\Sigma$. Using the fact that $\mathcal{C}^\infty\left(\left(a,b\right),\mathbb{C}\right)\otimes\mathcal{C}^\infty\left(\Sigma,\mathbb{C}^4\right)$ can be identified with a dense linear subspace of $\mathcal{C}^\infty\left(\left(a,b\right)\times\Sigma,\mathbb{C}^4\right)$ in a continuous way and similar $\mathcal{C}^\infty\left(\left(a,b\right),\mathbb{C}\right)\otimes\mathcal{C}^\infty\left(\Sigma,\left(\mathbb{C}^4\right)^*\right)$ can be continuously identified with a dense linear subspace of $\mathcal{C}^\infty\left(\left(a,b\right)\times\Sigma,\left(\mathbb{C}^4\right)^*\right)$, this all implies that the Dirac operator for spinors and cospinors can be written in split form:
\begin{flalign*}
&&D^\sp=-\iu\frac{\partial}{\partial t}\otimes\gamma^0+\bbone\otimes\gamma^0H^\sp&&\text{and}&&D^\cosp=\iu\frac{\partial}{\partial t}\otimes\gamma^0+\bbone\otimes H^\cosp\left(\cdot\right)\gamma^0,&&
\end{flalign*}
where $\bbone$ is the identity on $\mathcal{C}^\infty\left(\left(a,b\right),\mathbb{C}\right)$,
\begin{flalign}\label{spatial spinor operator}
&&H^\sp f&:=-\iu\gamma^0\gamma^i\left(\eta_i\left(f^A\right)e_A+\Gamma_i\left(t,\cdot\right)f\right)+m\gamma^0f,&&\makebox[0pt][r]{$f\in\mathcal{C}^\infty\left(\Sigma,\mathbb{C}^4\right)$,}
\intertext{and}\label{spatial cospinor operator}
&&H^\cosp\varphi&:=\iu\left(\eta_i\left(\varphi_A\right)e^{A}-\varphi\,\Gamma_i\left(t,\cdot\right)\right)\gamma^i\gamma^0+m\varphi\gamma^0,&&\makebox[0pt][r]{$\varphi\in\mathcal{C}^\infty\left(\Sigma,\left(\mathbb{C}^4\right)^*\right)$,}
\end{flalign}
with $e_A$ and $e^A$ as above. Recall that it does not matter which $t\in\left(a,b\right)$ is taken because of the time-independence of $\Gamma_i$.
Equation \eqref{spatial spinor operator} (resp., \eqref{spatial cospinor operator})
define $H^\sp$ (resp., $H^\cosp$) both as a partial
differential operator and also as an operator on the dense domain
$\mathcal{C}^\infty\left(\Sigma,\mathbb{C}^4\right)$ (resp., $\mathcal{C}^\infty\left(\Sigma,\left(\mathbb{C}^4\right)^*\right)$
in the Hilbert space $L^2\left(\Sigma,\mathbb{C}^4;\vol_h\right)$ (resp.,  $L^2\left(\Sigma,\left(\mathbb{C}^4\right)^*;\vol_h\right)$) 
with inner product
\begin{flalign*}
&&\langle\cdot\mid\cdot\rangle_2:\mathcal{C}^\infty\left(\Sigma,\mathbb{C}^4\right)\times\mathcal{C}^\infty\left(\Sigma,\mathbb{C}^4\right)\longrightarrow\mathbb{C},
&&\left(f,g\right)&\longmapsto\int\limits_\Sigma f^* g\vol_h,&&
\intertext{resp.,}
&&\langle\cdot\mid\cdot\rangle_2:\mathcal{C}^\infty\left(\Sigma,\left(\mathbb{C}^4\right)^*\right)\times\mathcal{C}^\infty\left(\Sigma,\left(\mathbb{C}^4\right)^*\right)\longrightarrow\mathbb{C},
&&\left(\varphi,\psi\right)&\longmapsto\int\limits_\Sigma\psi\,\varphi^*\vol_h,&&
\end{flalign*}
with the Hermitean conjugation applied pointwise in the integrands, 
so $f^* g$ and $\psi\varphi^*$ are smooth functions. 

%=====================================================================================================================================================================

%=====================================================================================================================================================================
%lemma
%=====================================================================================================================================================================
\begin{lem} \label{lem spatial Dirac operators are self-adjoint}
The partial differential operators $H^\sp$ (resp.,  $H^\cosp$) are elliptic, and define symmetric operators on
their domains of definition.
\end{lem}
%=====================================================================================================================================================================
\noindent\textbf{\textit{Proof}:} We only prove the ellipticity statement for $H^\cosp$, as the proof for $H^\sp$ is analogous.  From (\ref{spatial cospinor operator}), the principal symbol\footnote{For a short but yet insightful introduction to linear differential operators and their principal symbols (and the notations involved) see \cite[Sec.A.4]{BGP07} and \cite[Sec.1.2]{SWald12}.} of $H^\cosp$ is seen to be $\sigma_{H^\cosp}\left(\xi\right)=\iu\xi_i\gamma^i\gamma^0$ for $\xi\in T^*\Sigma$. One easily computes the determinant $\det\left(\xi_i\gamma^i\gamma^0\right)=\left(\xi_1^2+\xi_2^2+\xi_3^2\right)^2$, which shows that $\sigma_{H^\cosp}\left(\xi\right)$ is an isomorphism of complex vector spaces for all $\xi\in T^*\Sigma$ unless $\xi=0\in T^*\Sigma_x$ for $x\in\Sigma$.

Symmetry of the operators on the given domains (termed `self-adjointness'
in \cite[Chap.III, \S5]{LM89}) follows from
Stokes' theorem, given the easily proved identities
\begin{flalign*}
&&\left(H^\sp f\right)^*g-f^*\left(H^\sp g\right)=\iu d\left(f^*\gamma^0\gamma^ig\right)\left(\eta_i\right)&&\forall f,g\in\mathcal{C}^\infty\left(\Sigma,\mathbb{C}^4\right)
\intertext{and}
&&\psi\left(H^\cosp\varphi\right)^*-\left(H^\cosp\psi\right)\varphi^*=\iu d\left(\psi\,\gamma^0\gamma^i\varphi^*\right)\left(\eta_i\right)&&\forall\varphi,\psi\in\mathcal{C}^\infty\left(\Sigma,\left(\mathbb{C}^4\right)^*\right).
\end{flalign*}
\hfill\SquareCastShadowTopRight\par\bigskip
%=====================================================================================================================================================================
Given this result, we may apply \cite[Thm III.5.8]{LM89} to conclude that the eigenvalues of $H^\sp$ and $H^\cosp$ are real, have finite multiplicity, are countably many, say $\left\{\lambda_n\right\}_{n\in\mathbb{N}}$ and $\left\{\mu_n\right\}_{n\in\mathbb{N}}$, that these
sets of eigenvalues are unbounded in magnitude, and that their corresponding eigenfunctions are smooth.

Once normalised with respect to $\langle\cdot\mid\cdot\rangle_2$, we denote the smooth eigenfunctions by $\left\{\chi_n\in\mathcal{C}^\infty\left(\Sigma,\mathbb{C}^4\right)\right\}_{n\in\mathbb{N}}$ and $\left\{\zeta_n\in\mathcal{C}^\infty\left(\Sigma,\left(\mathbb{C}^4\right)^*\right)\right\}_{n\in\mathbb{N}}$ and their $L^2$-equivalence classes furnish orthonormal bases for $L^2\left(\Sigma,\mathbb{C}^4;\vol_h\right)$ and $L^2\left(\Sigma,\left(\mathbb{C}^4\right)^*;\vol_h\right)$. As we have the identities $\left(H^\sp  f\right)^\dagger=H^\cosp f^\dagger$ for all $f\in\mathcal{C}^\infty\left(\Sigma,\mathbb{C}^4\right)$ and $\left(H^\cosp\varphi\right)^\dagger=H^\sp\varphi^\dagger$ for all $\varphi\in\mathcal{C}^\infty\left(\Sigma,\left(\mathbb{C}^4\right)^*\right)$, 
$H^\sp$ and $H^\cosp$ have identical eigenvalues and, for all $\lambda\in\mathbb{R}$,
the $\lambda$-eigenspace of $H^\sp$ is mapped bijectively to the $\lambda$-eigenspace of $H^\cosp$ by 
the Dirac adjoint, now extended to an antiunitary map from $L^2\left(\Sigma,\mathbb{C}^4;\vol_h\right)$ to $L^2\left(\Sigma,\left(\mathbb{C}^4\right)^*;\vol_h\right)$ (we also refer to the inverse as the Dirac adjoint). 
Hence, without loss of generality we may assume $\lambda_n=\mu_n$ and $\chi_n^\dagger =\zeta_n$ for all $n\in\mathbb{N}$.
Indeed, we can do more: we may also assume that the 
the eigenvalues and smooth eigenfunctions of $H^\sp$ and $H^\cosp$ may be labelled
by the set $\mathbb{Z}':=\mathbb{Z}\setminus\left\{0\right\}$ so that: 
\begin{align}\label{eq:eigenordering}
\begin{aligned}
&&\left\{\lambda_z\right\}_{z\in\mathbb{Z}'}:&&\ldots\leq\lambda_{-3}\leq\lambda_{-2}\leq\lambda_{-1}\leq -m <0<  m \leq\lambda_1\leq\lambda_2\leq\lambda_3\leq\ldots&&\text{and}&&\lambda_{-z}=-\lambda_z;&& 
\end{aligned}\notag \\
\begin{aligned}
&&\left\{\chi_z\right\}_{z\in\mathbb{Z}'},\left\{\zeta_z\right\}_{z\in\mathbb{Z}'}:&&H^\sp\chi_z=\lambda_z\chi_z,&&H^\cosp\zeta_z=\lambda_z\zeta_z,&&\chi^\dagger _z=\zeta_z&&\text{and}&&\zeta_z^\dagger =\chi_z,&&
\end{aligned}
\end{align}
together with 
\begin{flalign} \label{eq:eigennorming}
&&\langle\chi_w\mid\gamma^0\chi_z\rangle_2=\langle\zeta_w\mid\zeta_z\gamma^0\rangle_2=
\begin{cases}
\frac{m}{\lambda_z}&\text{if}\enspace z=w\\
\sqrt{1-\frac{m^2}{\lambda_z^2}}&\text{if}\enspace z=-w\\ 
0&\text{if}\enspace z\neq\pm w
\end{cases}&&\makebox[0pt][r]{$\forall w,z\in\mathbb{Z}'$.}
\end{flalign}
(The last property will be useful in Section \ref{sec FR}.) These assumptions can be justified as follows: 
let $\chi,\chi'\in \mathcal{C}^\infty\left(\Sigma,\mathbb{C}^4\right)$ be normalised eigenfunctions of $H^\sp$
with eigenvalues $\lambda,\lambda'$. It follows from the identity 
$\left\{H^\sp,\gamma^0\right\}=2m$ and Lemma \ref{lem spatial Dirac operators are self-adjoint} that
\begin{align*}
\left(\lambda+\lambda'\right)\langle\chi\mid \gamma^0\chi'\rangle_2=2m\,\langle\chi\mid\chi'\rangle_2
\end{align*}
from which we may deduce that the eigenvalues of $H^\sp$ and $H^\cosp$ are all nonzero (as $m>0$) and also that
\begin{align*}
\langle\chi\mid\gamma^0\chi'\rangle_2=
\begin{cases}
\frac{m}{\lambda}\,\langle\chi\mid\chi'\rangle_2&\text{if}\enspace\lambda=\lambda'\\
0&\text{if}\enspace\lambda\neq\pm\lambda'
\end{cases}
\end{align*}
(this is not exhaustive as the case $\lambda=-\lambda'$ is left open). 
Furthermore, applying Cauchy--Schwarz, $m/|\lambda|\, \|\chi\|_2^2 = |\ip{\chi}{\gamma^0\chi}_2| \le \|\chi\|_2 \|\gamma^0\chi\|_2
=\|\chi\|_2^2$, so we see that the spectrum does not intersect the mass gap $(-m,m)$. 

Continuing with $\chi$ as above, define $\eta = \gamma^0\chi - \ip{\chi}{\gamma^0\chi}_2\chi=(\gamma^0-m/\lambda)\chi$. 
By construction, $\eta$ is $L^2$-orthogonal to $\chi$, $\ip{\chi}{\eta}_2 = 0$,  and so
 $\|\eta\|_2^2 =  \|\gamma^0\chi\|_2^2 - |\ip{\chi}{\gamma^0\chi}_2|^2=1-m^2/\lambda^2$ by Pythagoras' theorem.
Direct calculation now shows that $H^\sp\eta=-\lambda\eta$. Thus to every normalised eigenfunction $\chi$ with eigenvalue $\lambda\neq \pm m$, 
there is a normalised eigenfunction 
\[
\tilde{\eta}= (1-m^2/\lambda^2)^{-1/2}(\gamma^0-m/\lambda)\chi
\] 
with eigenvalue $-\lambda$. With this choice we also have 
\[
\ip{\tilde{\eta}}{\gamma^0\chi}_2=  \frac{1}{\sqrt{1-m^2/\lambda^2}} \left( \ip{\gamma^0\chi}{\gamma^0\chi}_2 - \frac{m}{\lambda} \ip{\chi}{\gamma^0\chi}_2\right) = \sqrt{1-m^2/\lambda^2}
\]
and further direct calculation shows that $(\gamma^0-m/\lambda)\chi$ and $(\gamma^0-m/\lambda)\chi'$ are orthogonal
if $\chi$ and $\chi'$ are orthogonal eigenfunctions. In the case $\lambda=\pm m$, the calculations above show that
$\eta=0$ and hence $\gamma^0\chi=\pm \chi$; it then holds that $\gamma^5\chi$ is a normalised eigenfunction of $H^\sp$ with
eigenvalue $\mp m$ and $\ip{\gamma^5\chi}{\gamma^0\chi}_2=0$ -- application of $\gamma^5$ is also unitary and
preserves orthogonality. Accordingly, a complete system of orthonormal eigenfunctions may be chosen obeying the conditions 
of \eqref{eq:eigenordering} and \eqref{eq:eigennorming}. 

%=====================================================================================================================================================================

%=====================================================================================================================================================================
\subsection{Solutions of the free Dirac equations and their Hilbert spaces}
%=====================================================================================================================================================================
We now use the results of Subsection \ref{subsec spatial Dirac operators} to solve the Dirac equations (\ref{Dirac equation spinor}) and (\ref{Dirac equation cospinor}) on an oriented and globally hyperbolic ultrastatic spacetime (or slab) $\textbf{M}$ with compact spatial section and spin connections as in Subsection \ref{subsec spatial Dirac operators}. We will also construct Hilbert spaces from the solutions thus obtained for an ensuing CAR-quantisation. In this connection, our interest lies in all smooth solutions with smooth Cauchy data (recall that $\Sigma$ is assumed to be compact). Thus, by \cite[Thm.3.5]{BG12}, any solution of interest can be written in terms of the advanced-minus-retarded Green operators $S:=S^\mathrm{adv}-S^\mathrm{ret}$ and $C:=C^\text{adv}-C^\text{ret}$, i.e. as $Su$, $u\in\mathcal{C}_0^\infty\left(M,\mathbb{C}^4\right)$, and $Cv$, $v\in\mathcal{C}_0^\infty\left(M,\left(\mathbb{C}^4\right)^*\right)$.

For each $f\in\mathcal{C}^\infty\left(M,\mathbb{C}^4\right)$ and $\varphi\in\mathcal{C}^\infty\left(M,\left(\mathbb{C}^4\right)^*\right)$, $f_t:=f\left(t,\cdot\right)$ and $\varphi_t:=\varphi\left(t,\cdot\right)$ are smooth $\mathbb{C}^4$-valued and $\left(\mathbb{C}^4\right)^*$-valued functions and square-integrable on $\Sigma$ with respect to $\vol_h$ for all $t\in\left(a,b\right)$. (We allow the possibilities $a=-\infty$ or $b=+\infty$.)  Hence, we have the $L^2$-expansions (valid in $L^2\left(\Sigma,\mathbb{C}^4;\vol_h\right)$ and $L^2\left(\Sigma,\left(\mathbb{C}^4\right)^*;\vol_h\right)$ respectively)
\begin{align}\label{L2-expansions}
&&f_t=\sum_{z\in\mathbb{Z}'}\langle\chi_z\mid f_t\rangle_2\,\chi_z&&\text{and}&&\varphi_t=\sum_{z\in\mathbb{Z}'}\langle\zeta_z\mid\varphi_t\rangle_2\,\zeta_z,&&
\end{align}
where $\langle\chi_z\mid f_t\rangle_2$ and $\langle\zeta_z\mid\varphi_t\rangle_2$ are smooth functions in $t$ with first derivatives\footnote{If $\varphi$ is smooth and $\Sigma$ is compact, $\lim_{h\rightarrow0}\frac{\varphi_{t+h}\left(x\right)-\varphi_t\left(x\right)}{h}=\frac{\partial\varphi}{\partial t}\left(t,x\right)$ uniformly in $x\in\Sigma$, for each $t\in\left(a,b\right)$; by compactness of $\Sigma$, $\varphi_t$ is differentiable
in the $L^2$-sense with derivative $\frac{\partial\varphi}{\partial t}(t,\cdot$) and one may iterate this argument to deduce $L^2$-smoothness. 
Continuity of the $L^2$-inner product gives $\frac{\partial}{\partial t}\langle\zeta_z\mid \varphi_t\rangle_2= 
\langle\zeta_z\mid\frac{\partial\varphi}{\partial t}\left(t,\cdot\right)\rangle_2$. In the same way, one shows $\frac{d}{dt}\langle\chi_z\mid f_t\rangle_2=\langle\chi_z\mid\frac{\partial f}{\partial t}\left(t,\cdot\right)\rangle_2$.} $\langle\chi_z\mid\frac{\partial f}{\partial t}\left(t,\cdot\right)\rangle_2$ and $\langle\zeta_z\mid\frac{\partial\varphi}{\partial t}\left(t,\cdot\right)\rangle_2$.

Now suppose $\psi\in\mathcal{C}^\infty\left(M,\mathbb{C}^4\right)$ and $\alpha\in\mathcal{C}^\infty\left(M,\left(\mathbb{C}^4\right)^*\right)$ are solutions of the inhomogeneous Dirac equations on $\textbf{M}$,
\begin{align*}
\left(-\iu\frac{\partial}{\partial t}\otimes\gamma^0+\bbone\otimes\gamma^0H^\sp\right)\psi=u&&\text{and}&&\left(\iu\frac{\partial}{\partial t}\otimes\gamma^0+\bbone\otimes H^\cosp\left(\cdot\right)\gamma^0\right)\alpha=v,
\end{align*}
where $u\in\mathcal{C}_0^\infty\left(M,\mathbb{C}^4\right)$ and $v\in\mathcal{C}_0^\infty\left(M,\left(\mathbb{C}^4\right)^*\right)$, then for each $t\in\left(a,b\right)$,
\begin{flalign*}
&&\frac{\partial\psi}{\partial t}\left(t,\cdot\right)+\iu H^\sp\psi_t=\iu\gamma^0u_t&&\text{and}
&&\frac{\partial\alpha}{\partial t}\left(t,\cdot\right)-\iu H^\cosp\alpha_t=-\iu v_t\gamma^0.&&
\end{flalign*}
Taking the $L^2$-inner product with $\chi_w$ and $\zeta_w$ and using Lemma \ref{lem spatial Dirac operators are self-adjoint}, we find
\begin{flalign}\label{differential equations Fourier coefficients spinors}
&&\frac{d}{dt}\langle\chi_w\mid\psi_t\rangle_2+\iu\lambda_w\,\langle\chi_w\mid\psi_t\rangle_2=\langle\chi_w\mid\iu\gamma^0u_t\rangle_2&&
\intertext{and}\label{differential equations Fourier coefficients cospinors}
&&\frac{d}{dt}\langle\zeta_w\mid\alpha_t\rangle_2-\iu\lambda_w\,\langle\zeta_w\mid\alpha_t\rangle_2=\langle\zeta_w\mid-\iu v_t\gamma^0\rangle_2&&\makebox[0pt][r]{$\forall w\in\mathbb{Z}'$.}
\end{flalign}
These are ordinary and inhomogeneous first order differential equations with constant coefficients for the Fourier coefficients of $\psi_t$ and $\alpha_t$. We find for the retarded and the advanced Green function and for the solutions of (\ref{differential equations Fourier coefficients spinors}) and (\ref{differential equations Fourier coefficients cospinors}) defined by them ($z\in\mathbb{Z}'$):
\begin{align*}
S^\text{ret}_z\left(t,t'\right)=\begin{cases}0&\text{if}\enspace a<t\leq t'<b\\e^{\iu\lambda_z\left(t'-t\right)}&\text{if}\enspace a<t'\leq t<b\end{cases},
&&S^\mathrm{adv}_z\left(t,t'\right)=\begin{cases}-e^{\iu\lambda_z\left(t'-t\right)}&\text{if}\enspace a<t\leq t'<b\\0&\text{if}\enspace a<t'\leq t<b\end{cases},
\end{align*}
\begin{align*}
\left(S^\text{ret}_z\langle\chi_z\mid\iu\gamma^0u_{t'}\rangle_2\right)\left(t\right)&=\langle\chi_z\mid\left(S^\text{ret}u\right)_t\rangle_2=\int\limits_a^te^{\iu\lambda_z\left(t'-t\right)}\langle\chi_z\mid\iu\gamma^0u_{t'}\rangle_2\,dt',\\
\left(S^\text{adv}_z\langle\chi_z\mid\iu\gamma^0u_{t'}\rangle_2\right)\left(t\right)&=\langle\chi_z\mid\left(S^\text{adv}u\right)_t\rangle_2=\int\limits_t^b-e^{\iu\lambda_z\left(t'-t\right)}\langle\chi_z\mid\iu\gamma^0u_{t'}\rangle_2\,dt'
\end{align*}
and $C^\text{ret/adv}_z=S^\text{ret/adv}_{-z}$. From this, we conclude for $u\in\mathcal{C}_0^\infty\left(M,\mathbb{C}^4\right)$, $v\in\mathcal{C}_0^\infty\left(M,\left(\mathbb{C}^4\right)^*\right)$ and $t\in\left(a,b\right)$ in the $L^2$-sense:
\begin{align}\label{spinor solution L2-expansion}
\left(Su\right)_t&=\sum_{z\in\mathbb{Z}'}\int\limits_a^b-e^{\iu\lambda_zt'}\langle\chi_z\mid\iu\gamma^0u_{t'}\rangle_2\,dt'\,e^{-\iu\lambda_zt}\chi_z
\intertext{and}\label{cospinor solution L2-expansion}
\left(Cv\right)_t&=\sum_{z\in\mathbb{Z}'}\int\limits_a^be^{-\iu\lambda_zt'}\langle\zeta_z\mid\iu v_{t'}\gamma^0\rangle_2\,dt'\,e^{\iu\lambda_zt}\zeta_z.
\end{align}
From this one can also see that $\left(Su\right)^\dagger=Cu^\dagger$ and $\left(Cv\right)^\dagger=Sv^\dagger$ for all $u\in\mathcal{C}^\infty_0\left(M,\mathbb{C}^4\right)$ and for all $v\in\mathcal{C}^\infty_0\left(M,\left(\mathbb{C}^4\right)^*\right)$.
In addition, one sees that
\begin{align}\label{spinor solution L2-norm}
\|\left(Su\right)_t\|^2&=\sum_{z\in\mathbb{Z}'} \left|
\int\limits_a^b e^{\iu\lambda_zt'}\langle\chi_z\mid\iu\gamma^0u_{t'}\rangle_2\,dt'
\right|^2
\end{align}
which is evidently constant in $t$; similar results apply to $\|\left(Cv\right)_t\|$. 

The cospinor solution space $\mathcal{L}^\cosp:=C\mathcal{C}_0^\infty\left(M,\left(\mathbb{C}^4\right)^*\right)$ becomes a pre-Hilbert space with the inner product $\langle\cdot\mid\cdot\rangle_\cosp$ (cf. \cite[Lem.4.2.4]{San08}):
\begin{flalign*}
&&\langle Cv\mid Cv'\,\rangle_\cosp
	=-\iu\int\limits_MCv'\,v^\dagger\vol_\textbf{M},&&\makebox[0pt][r]{$v,v'\in\mathcal{C}_0^\infty\left(M,\left(\mathbb{C}^4\right)^*\right)$,}
\end{flalign*}
while $\mathcal{L}^\sp:=S\mathcal{C}_0^\infty\left(M,\mathbb{C}^4\right)$ becomes a pre-Hilbert space with the inner product $\langle\cdot\mid\cdot\rangle_\sp$:
\begin{flalign}\label{spinor inner product}
&&\langle Su\mid Su'\,\rangle_\sp
	=\iu\int\limits_Mu^\dagger\,Su'\vol_\textbf{M},&&\makebox[0pt][r]{$u,u'\in\mathcal{C}_0^\infty\left(M,\mathbb{C}^4\right)$.}
\end{flalign}
The positivity of this inner product is established by the identity (e.g., \cite[Lem.4.2.4]{San08})
\begin{flalign}\label{eq:spL2}
&&\langle Su\mid Su\,\rangle_\sp
	= \|(Su)_t\|_2^2, && t\in (a,b),~u\in\mathcal{C}^\infty_0\left(M,\mathbb{C}^4\right).
\end{flalign}
%=====================================================================================================================================================================

%=====================================================================================================================================================================
%lemma
%=====================================================================================================================================================================
\begin{lem}\label{lem orthonormal bases} An orthonormal basis for $\left(\mathcal{L}^\emph{sp},\langle\cdot\mid\cdot\rangle_\emph{sp}\right)$ is given by $\left\{e^{-\iu\lambda_z\,\cdot}\,\chi_z\right\}_{z\in\mathbb{Z}'}$ and an orthonormal basis for $\left(\mathcal{L}^\emph{cosp},\langle\cdot\mid\cdot\rangle_\emph{cosp}\right)$ is $\left\{e^{\iu\lambda_z\,\cdot}\,\zeta_z\right\}_{z\in\mathbb{Z}'}$.
\end{lem}
%=====================================================================================================================================================================
\noindent\textbf{\textit{Proof}:} First of all, we show $e^{\iu\lambda_z\,\cdot}\,\zeta_z\in\mathcal{L}^\cosp$. To this end, let $\sigma\in\mathcal{C}^\infty_0\left(\left(a,b\right),\mathbb{R}\right)$ have unit integral and let $w\in\mathbb{Z}'$. Then $-\iu \sigma e^{\iu\lambda_w\,\cdot}\,\zeta_w\gamma^0$ has compact support and we find from (\ref{cospinor solution L2-expansion})
\begin{align*}
\left(C\left(-\iu\sigma e^{\iu\lambda_w\,\cdot}\,\zeta_w\gamma^0\right)\right)_t
&	=\sum_{z\in\mathbb{Z}'}\int\limits_a^be^{-\iu\lambda_zt'}\langle\zeta_z\mid\sigma\left(t'\right)e^{\iu\lambda_wt'}\zeta_w\rangle_2\,dt'\,e^{\iu\lambda_zt}\zeta_z
	=\int\limits_a^b\sigma\left(t'\right)dt'\,e^{\iu\lambda_wt}\zeta_w\\
&	=e^{\iu\lambda_wt}\zeta_w,
\end{align*}
where the equation is to be understood in the $L^2$-sense. Because a smooth representative of an $L^2$-equivalence class is unique, we obtain the result $C\left(-\iu\sigma e^{\iu\lambda_w\,\cdot}\,\zeta_w\gamma^0\right)=e^{\iu\lambda_w\,\cdot}\,\zeta_w$. 
Similarly, $S\left(\iu\sigma e^{-\iu\lambda_z\,\cdot}\,\gamma^0\chi_z\right)=e^{-\iu\lambda_z\,\cdot}\,\chi_z$. 
With these results, it is not difficult to prove that
$\left\{e^{-\iu\lambda_z\,\cdot}\,\chi_z\right\}_{z\in\mathbb{Z}'}$ 
(resp., $\left\{e^{\iu\lambda_z\,\cdot}\,\zeta_z\right\}_{z\in\mathbb{Z}'}$) are orthonormal systems in 
their appropriate spaces. We leave this to the reader and concentrate on completeness. Here, the simplest argument is to combine \eqref{eq:spL2} with \eqref{spinor solution L2-norm} to show that
\begin{flalign*}
\langle Su\mid Su\,\rangle_\sp
	= \sum_{z\in\mathbb{Z}'} \left|
\int\limits_a^b e^{\iu\lambda_zt'}\langle\chi_z\mid\iu\gamma^0u_{t'}\rangle_2\,dt'
\right|^2 = \sum_{z\in\mathbb{Z}'} \left| \ip{e^{-\iu\lambda_z\cdot}\chi_z}{Su}_\sp\right|^2,
\end{flalign*}
establishing completeness and concluding the proof. \hfill\SquareCastShadowTopRight\par\bigskip
%=====================================================================================================================================================================

%=====================================================================================================================================================================
%text
%=====================================================================================================================================================================
Consequently, the completions of $\mathcal{L}^\sp$ with respect to $\langle\cdot\mid\cdot\rangle_\sp$ and of $\mathcal{L}^\cosp$ with respect to $\langle\cdot\mid\cdot\rangle_\cosp$ yield Hilbert spaces $\left(\mathcal{H}^\sp,\langle\cdot\mid\cdot\rangle_\sp\right)$ and $\left(\mathcal{H}^\cosp,\langle\cdot\mid\cdot\rangle_\cosp\right)$ with orthonormal bases $\left\{e^{-\iu\lambda_z\,\cdot}\,\chi_z\right\}_{z\in\mathbb{Z}'}$ and $\left\{e^{\iu\lambda_z\,\cdot}\,\zeta_z\right\}_{z\in\mathbb{Z}'}$. In what follows, we will consider their direct Hilbert space sum $\mathcal{H}:=\mathcal{H}^\sp\oplus\mathcal{H}^\cosp$, where $\mathcal{H}^\cosp:=\overline{\mathcal{L}}\,\!^\sp$ and $\mathcal{H}^\cosp:=\overline{\mathcal{L}}\,\!^\cosp$ for the CAR-quantisation.
%=====================================================================================================================================================================

%=====================================================================================================================================================================
\subsection{CAR-quantisation and reference state}\label{subsec CAR-quantisation}
%=====================================================================================================================================================================
The Dirac adjoint $\dagger:\mathcal{C}^\infty\left(M,\mathbb{C}^4\right)\longrightarrow\mathcal{C}^\infty\left(M,\left(\mathbb{C}^4\right)^*\right)$ and its inverse (also denoted by $\dagger$) descend to well-defined antiunitary maps $\dagger:\mathcal{L}^\sp\longrightarrow\mathcal{L}^\cosp$ and $\dagger:\mathcal{L}^\cosp\longrightarrow\mathcal{L}^\sp$. Hence, we obtain an antiunitary involution
of $\mathcal{L}^\sp\oplus\mathcal{L}^\cosp$ by 
\begin{flalign*}
\left(\psi\oplus\alpha\right)^\dagger :=\alpha^\dagger\oplus \psi^\dagger
\end{flalign*}
Because of the involutive property, $\dagger$ is bounded with norm $\lVert\dagger \rVert=1$ and extends to $\mathcal{H}$. 

We may now form the self-dual CAR-algebra $\mathfrak{A}=\mathfrak{A}_\text{SDC}\left(\mathcal{H},\langle\cdot\mid\cdot\rangle,\dagger \right)$, which is the unital *-algebra generated by the elements of the form $B\left(\psi\oplus\alpha\right)$ and their conjugates $B\left(\psi\oplus\alpha\right)^*$, $\psi\oplus\alpha\in\mathcal{H}$, satisfying (see \cite[\S2]{Araki70}):
\begin{enumerate}
\item[(1)] Linearity: for all $\lambda,\mu\in\mathbb{C}$, for all $\psi\oplus\alpha,\varphi\oplus\beta\in\mathcal{H}$,
\begin{equation*}
B\left(\lambda\psi\oplus\alpha+\mu\varphi\oplus\beta\right)=\lambda B\left(\psi\oplus\alpha\right)+\mu B\left(\varphi\oplus\beta\right).
\end{equation*}
\item[(2)] CARs: for all $\psi\oplus\alpha,\varphi\oplus\beta\in\mathcal{H}$,
\begin{equation*}
B\left(\psi\oplus\alpha\right)B\left(\varphi\oplus\beta\right)^*+B\left(\varphi\oplus\beta\right)^*B\left(\psi\oplus\alpha\right)
=\langle\varphi\oplus\beta\mid\psi\oplus\alpha\rangle\cdot 1_\mathfrak{A}.
\end{equation*}
\item[(3)] Hermiticity: for all $\psi\oplus\alpha\in\mathcal{H}$, 
\begin{equation*}
B\left(\psi\oplus\alpha\right)^*=B\left(\left(\psi\oplus\alpha\right)^\dagger\right).
\end{equation*}
\end{enumerate}
$\mathfrak{A}$ has a unique $C$*-norm and we consider its completion $\overline{\mathfrak{A}}$ with respect to this norm. The smeared quantum Dirac spinor field is defined by $\Psi\left(v\right):=B\left(0\oplus Cv\right)$, $v\in\mathcal{C}^\infty_0\left(M,\left(\mathbb{C}^4\right)^*\right)$, and the smeared quantum Dirac cospinor field is $\Psi^\dagger\left(u\right):=B\left(Su\oplus0\right)$, $u\in\mathcal{C}^\infty_0\left(M,\mathbb{C}^4\right)$. The algebra $\overline{\mathfrak{A}}$ has a $U(1)$ global gauge group of unit-preserving $*$-automorphisms determined by $\eta_\lambda B\left(\psi\oplus\alpha\right) = B\left((\lambda\psi)\oplus(\overline{\lambda}\alpha)\right)$, where $\lambda\in\mathbb{C}$, $|\lambda|=1$.

We now construct our reference (Hadamard) state $\omega_0:\overline{\mathfrak{A}}\longrightarrow\mathbb{C}$. 
Introducing  
\begin{flalign*}
&&\kappa^\pm_z(t,x):= e^{-\iu\lambda_{\pm z}t}\chi_{\pm z}(x), &&z\in\mathbb{Z},
\end{flalign*}
so $\kappa^-_z = \kappa^+_{-z}$ for all $z\in\mathbb{Z}$, and the $\kappa^+_z$ (resp., $\kappa^-_z$) are positive (resp., negative)
frequency solutions for $z\in\mathbb{Z}^+$, we  define $Q^\sp:\mathcal{L}^\sp\longrightarrow\mathcal{L}^\sp$ to be the orthogonal projection onto the linear subspace of $\mathcal{L}^\sp$ which is spanned by $\left\{\kappa^+_z\mid z\in\mathbb{Z}^+\right\}$ (positive frequency spinor solutions), i.e. 
\begin{flalign*}
&&Q^\sp\psi
	=\sum_{z\in\mathbb{Z}^+}\langle \kappa^+_z\mid \psi\rangle_\sp\,\kappa^+_z,		&&\psi\in\mathcal{L}^\sp,
\end{flalign*}
where $\mathbb{Z}^+:=\left\{z\in\mathbb{Z}\mid z>0\right\}$, and extend continuously to $\mathcal{H}^\sp$. Similarly, we define $Q^\cosp:\mathcal{L}^\cosp\longrightarrow\mathcal{L}^\cosp$ to be the orthogonal projection onto the linear subspace of $\mathcal{L}^\cosp$ which is spanned by $\left\{e^{\iu\lambda_zt}\zeta_z\mid z\in\mathbb{Z}^-\right\}$ (positive frequency cospinor solutions), i.e.
\begin{flalign*}
&&Q^\cosp\alpha
=\sum_{z\in\mathbb{Z}^-}\langle \kappa^{+\dagger}_z \mid \psi\rangle_\cosp\,\kappa^{+\dagger}_z 
	=\sum_{z\in\mathbb{Z}^+}\langle \kappa^{-\dagger}_z \mid \psi\rangle_\cosp\,\kappa^{-\dagger}_z,		&&\psi\in\mathcal{L}^\sp,
\end{flalign*}
where $\mathbb{Z}^-:=\left\{z\in\mathbb{Z}\mid z<0\right\}$, and extend continuously to $\mathcal{H}^\cosp$. Observe the relations $Q^\sp=\id_{\mathcal{H}^\sp}-\dagger Q^\cosp\dagger$ and $Q^\cosp=\id_{\mathcal{H}^\cosp}-\dagger Q^\sp\dagger$. Then, $P:=Q^\sp\oplus Q^\cosp$ is a projection operator on $\mathcal{H}$ and so $0\leq P=P^*\leq1$. It is an easy exercise now to verify that $P+\dagger P\dagger =\id_\mathcal{H}$, thus $P$ meets (3.4) and (3.5) of \cite{Araki70}. Owing to \cite[Lem.3.3 \& Lem.4.3]{Araki70}, $P$ defines a gauge invariant \cite{PS70}, pure and quasifree state $\omega_0$ on $\overline{\mathfrak{A}}$ which is uniquely determined by \cite[(3.3)]{Araki70}, that is, 
\begin{flalign}\label{reference state}
&&\omega_0\left(B\left(\psi\oplus\alpha\right)^*B\left(\varphi\oplus\beta\right)\right)&=
\langle\psi\oplus\alpha\mid P(\varphi\oplus\beta)\rangle&&\forall\psi\oplus\alpha,\varphi\oplus\beta\in\mathcal{H}.\\
&&&=\langle\psi\mid Q^\sp\varphi\rangle_\sp + \langle\alpha\mid Q^\cosp\beta\rangle_\cosp &&\notag
\end{flalign}
Here, gauge invariance means that $\omega_0\circ\eta_\lambda=\omega_0$ for all $\lambda\in\mathbb{C}$, $|\lambda|=1$,
and is manifest from the preceding expression. 
The state $\omega_0$ is Hadamard \cite{SV00,DH06} and the associated Wightman two-point distribution can be written as
\begin{flalign}\label{reference 2-point distribution}
&&W^{\left(2\right)}_0&\left[\left(u\oplus v\right)\otimes\left(u'\oplus v'\right)\right]&&\\\nonumber
&&&:= \omega_0\left(B\left(Su\oplus Cv\right)B\left(Su'\oplus Cv'\right)\right) &&\\\nonumber
&&&= \omega_0\left(B\left(Sv^\dagger\oplus Cu^\dagger\right)^*B\left(Su'\oplus Cv'\right)\right)&&\\\nonumber
&&&=\langle Sv^\dagger\mid Q^\sp Su'\rangle_\sp+\langle Cu^\dagger\mid Q^\cosp Cv'\rangle_\cosp &&\displaybreak[0]\\\nonumber
&&&=\sum_{z\in\mathbb{Z}^+}\langle Sv^\dagger\mid \kappa_{z}^+\rangle_\sp \langle\kappa_{z}^+\mid  Su'\rangle_\sp
+\sum_{z\in\mathbb{Z}^+}\langle Cu^\dagger\mid \kappa_{z}^{-\dagger}\rangle_\cosp \langle\kappa_{z}^{-\dagger}\mid Cv'\rangle_\cosp
&&\\\nonumber
&&&=\sum_{z\in\mathbb{Z}^+} \int\limits_M\int\limits_M
\Big\{v\left(t,x\right)\kappa_{z}^+\left(t,x\right)
\kappa_{z}^+\left(t',x'\right)^\dagger u'\left(t',x'\right)\Big\}\vol_\textbf{M}\vol'_\textbf{M}
&&\displaybreak[0]\\\nonumber
&&&\phantom{=}+\sum_{z\in\mathbb{Z}^+}
\int\limits_M\int\limits_M
\Big\{\kappa_{z}^-\left(t,x\right)^\dagger u\left(t,x\right)
v'\left(t',x'\right)\kappa_{z}^-\left(t',x'\right)\Big\}\vol_\textbf{M}\vol'_\textbf{M}
&&\displaybreak[0]\\\nonumber
&&&&&\makebox[0pt][r]{$u,u'\in\mathcal{C}^\infty_0\left(M,\mathbb{C}^4\right)$, $v,v'\in\mathcal{C}^\infty_0\left(M,\left(\mathbb{C}^4\right)^*\right)$.}
\end{flalign}
In terms of the eigenfunctions $\chi_z$ and $\zeta_z$, this reads
\begin{flalign}
&&W^{\left(2\right)}_0&\left[\left(u\oplus v\right)\otimes\left(u'\oplus v'\right)\right]&&\\\nonumber
&&&=\sum_{z\in\mathbb{Z}^+}\int\limits_M\int\limits_Me^{-\iu\lambda_z\left(t-t'\right)}
		v\left(t,x\right)\chi_z\left(x\right)\zeta_z\left(x'\right)u'\left(t',x'\right)\vol_\textbf{M}\vol'_\textbf{M}&&\\\nonumber
&&&	\phantom{=}+\sum_{z\in\mathbb{Z}^-}\int\limits_M\int\limits_Me^{\iu\lambda_z\left(t-t'\right)}
		\zeta_z\left(x\right)u\left(t,x\right)v'\left(t',x'\right)\chi_z\left(x'\right)\vol_\textbf{M}\vol'_\textbf{M},&&\\\nonumber
&&&&&\makebox[0pt][r]{$u,u'\in\mathcal{C}^\infty_0\left(M,\mathbb{C}^4\right)$, $v,v'\in\mathcal{C}^\infty_0\left(M,\left(\mathbb{C}^4\right)^*\right)$.}
\end{flalign}
Using the reference state, we can 
determine whether or not any other state is Hadamard, by whether or not the difference of their
Wightman two-point distributions is smooth.  For this purpose, 
it is useful to use the fact that $Q^\cosp=\id_{\mathcal{H}^\cosp}-\dagger Q^\sp\dagger$ to write
\begin{flalign}\label{eq:W20sp}
 W^{\left(2\right)}_0 \left[\left(u\oplus v\right)\otimes\left(u'\oplus v'\right)\right] =\langle Sv^\dagger\mid Q^\sp Su'\rangle_\sp -
\langle Sv^{\prime\dagger}\mid Q^\sp Su\rangle_\sp
+\langle Cu^\dagger\mid  Cv'\rangle_\cosp.
\end{flalign}

%=====================================================================================================================================================================

%=====================================================================================================================================================================
\section{FP-states on globally hyperbolic ultrastatic slabs with compact spatial sections}\label{sec FR}
%=====================================================================================================================================================================
From now on, we focus on the situation for oriented and globally hyperbolic ultrastatic slabs with compact spatial section and spin connections as constructed in Subsection \ref{subsec spatial Dirac operators}, i.e. $a,b\in\mathbb{R}$ are now taken such that $-\infty<a<b<\infty$. In this section, we will show how the fermionic projector description of \cite{FR14a} gives rise to a gauge invariant, pure and quasifree state on the $C$*-completion $\overline{\mathfrak{A}}$ of the self-dual CAR-algebra $\mathfrak{A}=\mathfrak{A}_\text{SDC}\left(\mathcal{H},\langle\cdot\mid\cdot\rangle,\dagger \right)$ for the quantised free Dirac field (see Subsection \ref{subsec CAR-quantisation}) on such a spacetime $\textbf{M}$. We will call this state the \emph{unsoftened FP-state}. 
Proceeding in the spirit of \cite{BF14}, we 
regard the unsoftened FP-state as a special case within a class of FP-states,
parameterised by a nonnegative
integrable function $f\in L^1(\mathbb{R})$. Within this class, 
states obtained from smooth compactly supported $f$ will be described
as {\em softened FP-states}, while the unsoftened FP-state corresponds
to taking $f$ as the characteristic function of $(a,b)$.  Our main objectives
are to show that the unsoftened FP-state does not have the Hadamard property in general, but that the softened FP-states are all Hadamard. 

To begin, let $\textbf{N}$ be the oriented and globally hyperbolic ultrastatic spacetime with exactly the same compact spatial section and spin connections as $\textbf{M}$ [but with underlying manifold $N=\mathbb{R}\times\Sigma$]. By extension with zero, any $u\in\mathcal{C}^\infty_0\left(M,\mathbb{C}^4\right)$ can also be regarded as a smooth and compactly supported $\mathbb{C}^4$-valued function on $N$. In this regard, $\tilde{\psi}=S_\textbf{N}u\in\mathcal{L}^\sp_\textbf{N}$ constitutes the unique solution of (\ref{Dirac equation spinor}) on $N$ which coincides with the unique solution of (\ref{Dirac equation spinor}) on $M$, $\psi=Su\in\mathcal{L}^\sp$. 
Formulae or objects relating to $\textbf{N}$ will be denoted using a subscript `$\textbf{N}$'; otherwise $\textbf{M}$ is to be understood. 
 
Let $f\in L^1(\mathbb{R})$ be an integrable nonnegative function. Typically we will have in mind that $f$ is either 
the characteristic function $\chi_{\left(a,b\right)}$ of $\left(a,b\right)$ (which will yield the unsoftened FP-state) or 
a compactly supported smooth function (which will yield the softened FP-states). 
As $f$ is integrable, it has a Fourier transform, for which we adopt the nonstandard convention
\[
\hat{f}\left(\lambda\right):=\int_\mathbb{R}fe^{\iu \lambda t}dt\,.
\]
Essential to our construction is the non-degenerate Hermitean sesquilinear form
\begin{flalign*}
&&{<\cdot\mid\cdot>_{\text{FP}_f}}:\mathcal{L}^\sp\times\mathcal{L}^\sp\longrightarrow\mathbb{C},&&\left(\psi,\varphi\right)\longmapsto\int\limits_Nf\,\tilde{\psi}^\dagger\,\tilde{\varphi}\vol_\textbf{N},&&
\end{flalign*}
which reduces to the form studied in \cite{FR14a}, in the case where
$f$ is the characteristic function $\chi_{\left(a,b\right)}$ of $\left(a,b\right)$. 
Using the Cauchy-Schwarz inequality and (\ref{spinor solution L2-expansion}),
\begin{flalign*}
&&\lvert<\psi\mid\varphi>_{\text{FP}_f}\rvert&\leq\int\limits_\mathbb{R}f\,\lvert\langle\tilde{\psi}_t\mid\gamma^0\tilde{\varphi_t}\rangle_2\rvert\,dt\leq\int\limits_\mathbb{R}f\,\lVert\tilde{\psi}_t\rVert_2\,\lVert\tilde{\varphi}_t\rVert_2\,dt=\int\limits_\mathbb{R}f\,\lVert\tilde{\psi}\rVert_\textbf{N}^\sp\,\lVert\tilde{\varphi}\rVert^\sp_\textbf{N}\,dt&&\\
&&&\leq\hat{f}\left(0\right)\lVert\psi\rVert_\sp\,\lVert\varphi\rVert_\sp&&\makebox[0pt][r]{$\forall\psi,\varphi\in\mathcal{L}^\sp$,}
\end{flalign*}
where we have used \eqref{eq:spL2} to give   $\lVert\tilde{\psi}_t\rVert_2=\lVert\tilde{\psi}\rVert_\textbf{N}^\sp$.
Thus, ${<\cdot\mid\cdot>}_{\text{FP}_f}$ is continuous and by \cite[Thm.20.2.1]{BB03}, there is a unique self-adjoint bounded operator $A_f:\mathcal{H}^\sp\longrightarrow\mathcal{H}^\sp$ satisfying the identity $\langle\psi\mid A_f\varphi\rangle_\sp={<\psi\mid\varphi>_{\text{FP}_f}}$ for all $\psi,\varphi\in\mathcal{L}^\sp$. 

Our construction proceeds by defining $Q_f^\sp$ to be the spectral projection $\chi_{\mathbb{R}^+}(A_f)$
and using this to construct a state $\omega_{\text{FP}_f}$ by analogy with the construction of 
the reference state $\omega_0$ in terms of $Q^\sp$. 
This state will be known as the ${\text{FP}_f}$-state; if $f$ is the characteristic function of the interval $(a,b)$, 
we refer to $\omega_{\text{FP}_f}$ as the unsoftened FP-state for the ultrastatic slab $\textbf{M}$, while if
$f$ is smooth and compactly supported, we describe $\omega_{\text{FP}_f}$ as a softened FP-state. 
The definition of $Q^\sp_f$ is suggested by the 
constructions of~\cite{FR14a}, although we emphasise that
$Q^\sp_f$ is not itself the fermionic projector, which is 
closely related to the complementary spectral projection $\chi_{\mathbb{R}^-}(A_f)$.

The first step is to compute the action of $A_f$ on elements of $\mathcal{H}^\sp$. We realise that 
\begin{flalign*}
&&\langle \kappa^+_w\mid A_f\kappa^+_z\rangle_\sp
&={ <\kappa^+_w\mid \kappa^+_z>}_{\mathrm{FP}_f}
=\int\limits_N f e^{\iu\lambda_wt}e^{-\iu\lambda_zt}\chi_w^\dagger\,\chi_z\vol_\textbf{N}&&\displaybreak[0]\\
&&&=\int\limits_\mathbb{R}f e^{\iu\left(\lambda_w-\lambda_z\right)t}\,\langle\chi_w\mid\gamma^0\chi_z\rangle_2\,dt&&\displaybreak[0]\\
&&&=\frac{\hat{f}\left(0\right)m}{\lambda_z}\,\delta_{wz}+\hat{f}\left(2\lambda_z\right)\sqrt{1-\frac{m^2}{\lambda^2_z}}\,\delta_{-wz}&&\makebox[0pt][r]{$\forall w,z\in\mathbb{Z}'$,}
\end{flalign*}
and thus
\begin{flalign*}
&&\langle \kappa^+_z\mid A_f\psi\rangle_\sp
=\frac{\hat{f}\left(0\right)m}{\lambda_z}\,\langle \kappa^+_z\mid\psi\rangle_\sp+\hat{f}\left(2\lambda_z\right)\sqrt{1-\frac{m^2}{\lambda^2_z}}\,\langle \kappa^-_z\mid\psi\rangle_\sp, &&z\in\mathbb{Z}',~\psi\in\mathcal{H}^\sp.
\end{flalign*} 
We hence see that $A_f$ acts on $\mathcal{H}^\sp$ by 
\begin{flalign}\label{eq:Af}
A_f\psi=\sum_{z\in\mathbb{Z}'}\left(\frac{\hat{f}\left(0\right)m}{\lambda_z}\,\langle \kappa^+_z\mid\psi\rangle_\sp
		+\hat{f}\left(2\lambda_z\right)\sqrt{\!1-\!\frac{m^2}{\lambda_z^2}}\,\langle \kappa^-_z\mid\psi\rangle_\sp\right)\kappa^+_z .
\end{flalign}
In order to proceed with the description given in \cite{FR14a}, we need to obtain the spectral decomposition of $A_f$.
For this purpose, it is useful to note that $A_f$ evidently decomposes as a direct sum $A_f=\bigoplus_{z\in\mathbb{Z}^+} A_{f,z}$
with respect to the decomposition $\mathcal{H}^\sp =\bigoplus_{z\in\mathbb{Z}^+} \mathcal{H}_z^\sp$, where
$\mathcal{H}_z^\sp$ is the two-dimensional space spanned by $\kappa^\pm_z$.
Representing $\alpha \kappa^+_z+ \beta \kappa^-_z$ by the column
vector $\begin{pmatrix} \alpha & \beta\end{pmatrix}^\top$, $A_{f,z}$ takes matrix form
\begin{flalign}
A_{f,z} = \begin{pmatrix} 
\hat{f}\left(0\right)m/\lambda_z & \hat{f}\left(2\lambda_z\right)\sqrt{\!1-\!m^2/\lambda_z^2}
\\ \overline{\hat{f}\left(2\lambda_z\right)}\sqrt{\!1-\!m^2/\lambda_z^2}
& -\hat{f}\left(0\right)m/\lambda_z 
\end{pmatrix}
\end{flalign}
where we have used $\lambda_{-z}=-\lambda_z$ and also the fact that $f$ is real-valued. 
It is convenient to parameterise
\begin{flalign}\label{eq:param}
A_{f,z} =\Xi_{f,z} \begin{pmatrix} 
\cos 2\theta_{f,z} & e^{\iu \phi_{f,z}} \sin 2\theta_{f,z} \\
e^{-\iu \phi_{f,z}} \sin 2\theta_{f,z} & -\cos 2\theta_{f,z}
\end{pmatrix}
\end{flalign} 
where 
\begin{flalign}\label{eq:Xifz}
\Xi_{f,z}:= \sqrt{\left\lvert\hat{f}\left(2\lambda_z\right)\right\rvert^2\left(1-\frac{m^2}{\lambda_z^2}\right)+\frac{\hat{f}\left(0\right)^2m^2}{\lambda^2_z}}
\end{flalign}
and $\theta_{f,z}\in[-\pi/2,\pi/2)$ and $\phi_{f,z}\in [0,2\pi)$ are uniquely determined. Note that
$\Xi_{f,z}>0$ for each $z\in\mathbb{Z}^+$; hence $\cos 2\theta_{f,z}>0$ for $z\in\mathbb{Z}^+$
(as $f$ is nonnegative) and so $\cos\theta_{f,z}>1/\sqrt{2}$.
%=====================================================================================================================================================================

%=====================================================================================================================================================================
%lemma
%=====================================================================================================================================================================
\begin{lem}\label{lem spectrum of A} The spectrum of $A_f$ is a pure point spectrum, with eigenvalues
\begin{align}\label{eigenvalues A}
\sigma_p\left(A_f\right)=\left\{\pm\Xi_{f,z} \mid  z\in\mathbb{Z}^+\right\}.
\end{align}
A basis of normalised eigenvectors $\kappa^{\pm}_{f,z}$ ($z\in \mathbb{Z}^+$) is defined by 
\begin{align} 
\kappa^{+}_{f,z} &= \cos \theta_{f,z} \kappa^+_z +e^{-\iu \phi_{f,z}} \sin\theta_{f,z} \kappa^-_z\\
\kappa^{-}_{f,z} &=\cos \theta_{f,z}  \kappa^-_z -e^{\iu \phi_{f,z}}\sin\theta_{f,z} \kappa^+_z ,
\end{align}
so that $A_f \kappa^{\pm}_{f,z} = \Xi_{f,z} \kappa^{\pm}_{f,z}$ for each $z\in\mathbb{Z}^+$. 
\end{lem}
%=====================================================================================================================================================================
\noindent\textbf{\textit{Proof}:} Elementary computation (e.g., using the matrix form of $A_{f,z}$) shows that the
$\kappa^\pm_{f,z}$ are normalised eigenvectors of $A_f$ with the stated eigenvalues. As  $\kappa^\pm_{f,z}$ span $\mathcal{H}^\sp_z$
for each $z\in\mathbb{Z}^+$, they provide a complete orthonormal basis for $\mathcal{H}^\sp$ which demonstrates that the
spectrum is pure point. 
\hfill\SquareCastShadowTopRight\par\bigskip
 
%=====================================================================================================================================================================

%=====================================================================================================================================================================
%text
%=====================================================================================================================================================================
With the eigenvalues and the eigenvectors found in Lemma \ref{lem spectrum of A}, we have the spectral decomposition of $A_f$:
\begin{align*}
A_f= \sum_{z\in\mathbb{Z}^{+}, s=\pm}s\,\Xi_{f,z}\,\langle\kappa^s_{f,z}\mid\cdot\rangle_\sp\,\kappa^s_{f,z}
\end{align*} 
Proceeding by analogy with the fermionic projector prescription of \cite{FR14a},
we define $Q^\sp_{f}=\chi_{\left(0,\infty\right)}\left(A_f\right)$, the projection onto the positive eigenspace, given by
\begin{equation*}
Q^\sp_f=\sum_{z\in\mathbb{Z}^+}\langle\kappa^+_{f,z}\mid\cdot\rangle_\sp\,\kappa^+_{f,z}.
\end{equation*}
Note that, if $f=\chi_{\left(a,b\right)}$ then $\kappa^+_{f,z}\to e^{-\iu\lambda_z\cdot}\chi_z$ in the limit 
$a\rightarrow-\infty,\,b\rightarrow\infty$, for $z\in\mathbb{Z}^+$; one may show that $Q_f^\sp\to Q^\sp$ strongly in this limit, 
recovering the projection that defined the reference state. 

For future reference, it is also useful to note that $Q_f^\sp$ and $Q^\sp$ both decompose as a direct sums
with respect to the decomposition $\mathcal{H}^\sp =\bigoplus_{z\in\mathbb{Z}^+} \mathcal{H}_z^\sp$.
The components $Q_{f,z}^\sp$ and $Q_z^\sp$ corresponding to $\mathcal{H}_z^\sp$ have the matrix forms
\begin{flalign}
Q_{f,z}^\sp = \begin{pmatrix} \cos^2 \theta_{f,z} & e^{\iu \phi_{f,z}}\sin \theta_{f,z} \cos \theta_{f,z}\\ 
e^{-\iu \phi_{f,z}}\sin \theta_{f,z} \cos \theta_{f,z} & \sin^2 \theta_{f,z}
\end{pmatrix}, \qquad
Q_{z}^\sp = \begin{pmatrix} 1 & 0\\ 
0 & 0
\end{pmatrix}
\end{flalign}
so
\begin{flalign}\label{eq:Qdiff}
Q_{f,z}^\sp -Q_z^\sp = \begin{pmatrix} -\sin^2 \theta_{f,z} & e^{\iu \phi_{f,z}}\sin \theta_{f,z} \cos \theta_{f,z}\\ 
e^{-\iu \phi_{f,z}}\sin \theta_{f,z} \cos \theta_{f,z} & \sin^2 \theta_{f,z}
\end{pmatrix}.
\end{flalign}

In order to construct a gauge invariant, pure and quasifree state $\omega_{\text{FP}_f}$ on $\overline{\mathfrak{A}}$, we need to `double' $Q^\sp_f$ to a self-adjoint bounded operator $P_{\text{FP}_f}$ on $\mathcal{H}$ satisfying the two conditions \cite[Eqs.~(3.4), (3.5)]{Araki70}, that is, $0\leq P^*_{\text{FP}_f}=P_{\text{FP}_f}\leq 1$ and $P_{\text{FP}_f}+\dagger P_{\text{FP}_f}\dagger=\id_\mathcal{H}$. Consider the projection operator on $\mathcal{H}^\cosp$ defined by 
\begin{align*}
Q^\cosp_f
	:=\id_{\mathcal{H}^\cosp}-\dagger Q^\sp_f\dagger 
	=\sum_{z\in\mathbb{Z}^+}\left\langle \kappa^{-\dagger}_{f,z} \,\Big|\,\,\cdot\,\right\rangle_\cosp \kappa^{-\dagger}_{f,z} ,
\end{align*}
then also $Q^\sp_f=\id_{\mathcal{H}^\sp}-\dagger Q^\cosp_f\dagger$, and $P_{\text{FP}_f}=Q^\sp_f\oplus Q^\cosp_f:\mathcal{H}\longrightarrow\mathcal{H}$ has the required properties. Now, \cite[Lem.3.3 \& Lem.4.3]{Araki70} yield a gauge invariant, pure and quasifree state $\omega_{\text{FP}_f}:\overline{\mathfrak{A}}\longrightarrow\mathbb{C}$, the $\text{FP}_f$-state, uniquely determined by
\begin{flalign*}
&&\omega_{\text{FP}_f}\left(B\left(\psi\oplus\alpha\right)^*B\left(\varphi\oplus\beta\right)\right)=\langle\psi\oplus\alpha\mid P_{\text{FP}_f}\varphi\oplus\beta\rangle&&\forall\psi\oplus\alpha,\varphi\oplus\beta\in\mathcal{H}.
\end{flalign*}
The associated Wightman two-point distribution can thus be written as
\begin{flalign}\label{FR 2-point distribution}
&&W^{\left(2\right)}_{\text{FP}_f}&\left[\left(u\oplus v\right)\otimes\left(u'\oplus v'\right)\right]&&\displaybreak[0]\\\nonumber
&&&=\langle S v^\dagger\mid Q^\sp_fS u'\rangle_\sp+\langle C u^\dagger\mid Q^\cosp_f C v'\rangle_\cosp &&\displaybreak[0]\\\nonumber
&&&=\sum_{z\in\mathbb{Z}^+} \int\limits_M\int\limits_M
\Big\{v\left(t,x\right)\kappa_{f,z}^+\left(t,x\right)
\kappa_{f,z}^+\left(t',x'\right)^\dagger u'\left(t',x'\right)\Big\}\vol_\textbf{M}\vol'_\textbf{M}
&&\displaybreak[0]\\\nonumber
&&&\phantom{=}+\sum_{z\in\mathbb{Z}^+}
\int\limits_M\int\limits_M
\Big\{\kappa_{f,z}^-\left(t,x\right)^\dagger u\left(t,x\right)
v'\left(t',x'\right)\kappa_{f,z}^-\left(t',x'\right)\Big\}\vol_\textbf{M}\vol'_\textbf{M}
&&\displaybreak[0]\\\nonumber
&&&&&\makebox[0pt][r]{$u,u'\in\mathcal{C}^\infty_0\left(M,\mathbb{C}^4\right)$, $v,v'\in\mathcal{C}^\infty_0\left(M,\left(\mathbb{C}^4\right)^*\right)$.}
\end{flalign}
As in the case of the reference state, we may also write
\begin{flalign}\label{eq:W2fsp}
 W^{\left(2\right)}_{\text{FP}_f} \left[\left(u\oplus v\right)\otimes\left(u'\oplus v'\right)\right] =\langle Sv^\dagger\mid Q_f^\sp Su'\rangle_\sp -
\langle Sv^{\prime\dagger}\mid Q_f^\sp Su\rangle_\sp
+\langle Cu^\dagger\mid  Cv'\rangle_\cosp.
\end{flalign}

%=====================================================================================================================================================================
\subsection{The unsoftened FP-state is not Hadamard}\label{subsec FR-state is not Hadamard}
%=====================================================================================================================================================================
We can now establish that the unsoftened FP-state cannot be a Hadamard state in general, proceeding along similar lines to \cite{FeV12b}. As mentioned at the end of Section~\ref{subsec CAR-quantisation}, we can determine whether or not a state is Hadamard
by whether or not its two-point distribution differs from that of the reference state by a distribution that amounts to integration against 
a smooth function.   
At this point, it is requisite to specify what is to be understood exactly by ``\textit{integration against a smooth function}'': to be precise, $\omega_{\text{FP}_f}$ 
is Hadamard if and only if there exists a smooth function $k\in\mathcal{C}^\infty\left(M\times M,\left[\mathbb{C}^4\oplus\left(\mathbb{C}^4\right)^*\right]\otimes\left[\mathbb{C}^4\oplus\left(\mathbb{C}^4\right)^*\right]\right)$ such that
\begin{flalign}\label{eq:smoothness}
&&\left(W^{\left(2\right)}_{\text{FP}_f}-W^{\left(2\right)}_0\right)\left[\sigma\right]=\int\limits_{M\times M}k^*\sigma\vol_{\textbf{M}\times\textbf{M}}&&\\
&&&&\makebox[0pt][r]{$\forall\sigma\in\mathcal{C}_0^\infty\left(M\times M,\left[\mathbb{C}^4\oplus\left(\mathbb{C}^4\right)^*\right]\otimes\left[\mathbb{C}^4\oplus\left(\mathbb{C}^4\right)^*\right]\right)$,}\nonumber
\end{flalign}
where ``\,$*$\,'' denotes Hermitean conjugation as before. 

Since $\mathcal{C}_0^\infty\left(M,\mathbb{C}^4\oplus\left(\mathbb{C}^4\right)^*\right)\otimes\mathcal{C}_0^\infty\left(M,\mathbb{C}^4\oplus\left(\mathbb{C}^4\right)^*\right)$ can be identified in a continuous manner with a dense linear subspace of $\mathcal{C}_0^\infty\left(M\times M,\left[\mathbb{C}^4\oplus\left(\mathbb{C}^4\right)^*\right]\otimes\left[\mathbb{C}^4\oplus\left(\mathbb{C}^4\right)^*\right]\right)$, it is enough to establish \eqref{eq:smoothness} for $\sigma$ of the form $\sigma =  \left(u\oplus v\right)\otimes\left(u'\oplus v'\right)$ for $u,u'\in\mathcal{C}^\infty_0\left(M,\mathbb{C}^4\right)$ and $v,v'\in\mathcal{C}^\infty_0\left(M,\left(\mathbb{C}^4\right)^*\right)$. 
In fact, we need not consider the full difference $W^{\left(2\right)}_{\text{FP}_f}-W^{\left(2\right)}_0$ but only `half' of it. Comparing \eqref{eq:W20sp} and \eqref{eq:W2fsp}, we see that
\begin{multline*}
\left(W^{\left(2\right)}_{\text{FP}_f}-W^{\left(2\right)}_0\right)\left[\left(u\oplus v\right)\otimes\left(u'\oplus v'\right)\right] 
=\langle S v^\dagger\mid\left(Q^\sp_f-Q^\sp\right)S u'\rangle_\sp-\langle Sv'^\dagger\mid\left(Q^\sp_f-Q^\sp\right)Su\rangle_\sp\\
 u,u'\in\mathcal{C}^\infty_0\left(M,\mathbb{C}^4\right),\,  v,v'\in\mathcal{C}^\infty_0\left(M,\left(\mathbb{C}^4\right)^*\right) 
\end{multline*}
and as the two summands are of the same form, we conclude
that $\omega_{\text{FP}_f}$ is Hadamard if and only if there exists $k\in\mathcal{C}^\infty\left(M\times M,\mathbb{C}^4\otimes\left(\mathbb{C}^4\right)^*\right)$ such that
\[
\langle S v^\dagger\mid\left(Q^\sp_f-Q^\sp\right)S u'\rangle_\sp=\int\limits_{M\times M}k^*\left(u'\otimes v\right)\vol_{\textbf{M}\times\textbf{M}}
\]
Now, if such a smooth function $k$ exists, it is clearly smooth and square-integrable on the smooth product manifold $M'\times M'$
with respect to the product measure $\vol_{\textbf{M}'\times\textbf{M}'}$, where we have defined $M':=\left(a',b'\right)\times\Sigma$
for some choice of $a'$ and $b'$ with $a<a'<b'<b$. Thus $k$ defines a Hilbert--Schmidt operator $K$ on $L^2\left(M',\mathbb{C}^4;\vol_{\textbf{M}'}\right)$ such that
\begin{align*}
\langle\varphi\mid K\psi\rangle_{\left(a',b'\right)}& :=\int\limits_{M'\times M'}k^*\left(\gamma^0\psi\otimes\varphi^\dagger\gamma^0\right)\vol_{\textbf{M}'\times\textbf{M}'},
\qquad\qquad \varphi,\psi\in L^2\left(M',\mathbb{C}^4;\vol_{\textbf{M}'}\right),
\end{align*}
Now, as
\begin{align}\label{difference two-point distributions}
\langle S v^\dagger\mid\left(Q^\sp_f-Q^\sp\right)S u'\rangle_\sp &= \sum_{z\in\mathbb{Z}^+} \int\limits_M\int\limits_M
\Big\{v\left(t,x\right) \kappa_{f,z}^+\left(t,x\right)
\kappa_{f,z}^+\left(t',x'\right)^\dagger  
u'\left(t',x'\right)\Big\}\vol_\textbf{M}\vol'_\textbf{M}\nonumber \\
&\qquad - \sum_{z\in\mathbb{Z}^+} \int\limits_M\int\limits_M
\Big\{v\left(t,x\right) \kappa_{z}^+\left(t,x\right)
\kappa_{z}^+\left(t',x'\right)^\dagger  
u'\left(t',x'\right)\Big\}\vol_\textbf{M}\vol'_\textbf{M},
\end{align}
it follows that 
\begin{align*}
\langle\varphi\mid K\psi\rangle_{M'} &=
\sum_{z\in\mathbb{Z}^+} \left( \ip{\varphi}{\kappa^+_{f,z}}_{M'}\ip{\kappa^+_{f,z}}{\psi}_{M'} - 
\ip{\varphi}{\kappa^+_{z}}_{M'}\ip{\kappa^+_{z}}{\psi}_{M'}\right)
\end{align*}
where $\ip{\cdot}{\cdot}_{M'}$ is the inner product of  $L^2\left(M',\mathbb{C}^4;\vol_{\textbf{M}'}\right)$ 
(in the first instance, this formula is obtained for $\varphi,\psi\in \mathcal{C}^\infty_0\left(M,\mathbb{C}^4\right)$,
but one then extends by continuity). One sees that the span of $\kappa^\pm_{z}$ in $L^2\left(M',\mathbb{C}^4;\vol_{\textbf{M}'}\right)$
is an invariant subspace for $K$, for each $z\in\mathbb{Z}^+$. Writing the restriction of $K$ to this subspace as $K_z$, 
it is clear that $K_z$ has matrix form (cf.~\eqref{eq:Qdiff})
\[
K_z = (b'-a') \begin{pmatrix} -\sin^2 \theta_{f,z} & e^{\iu \phi_{f,z}}\sin \theta_{f,z} \cos \theta_{f,z}\\ 
e^{-\iu \phi_{f,z}}\sin \theta_{f,z} \cos \theta_{f,z} & \sin^2 \theta_{f,z}
\end{pmatrix}
\]
with respect to the orthonormal basis $\kappa^+_{z}/\sqrt{b'-a'}, \kappa^-_{z}/\sqrt{b'-a'}$. 
As this matrix is trace-free, it has a pair of eigenvalues with equal magnitude but opposite signs, where the 
magnitude is the square-root of the minus the determinant. Thus we deduce:
%=====================================================================================================================================================================
%lemma
%=====================================================================================================================================================================
\begin{lem}\label{lem eigenvalues of K*} The operator $K$ has spectrum
\begin{equation}\label{eigenvalue of K*}
\sigma(K) = \{0\} \cup \{ \pm(b'-a')\sin\theta_{f,z}\mid z\in\mathbb{Z}^+\} .
\end{equation}
\end{lem}
%=====================================================================================================================================================================
\noindent\textbf{\textit{Proof}:} Elementary calculation shows that $K_z$ has eigenvalues $\pm(b'-a')\sin\theta_{f,z}$,
with corresponding eigenvectors that must span the two-dimensional space spanned by $\kappa^\pm_{z}/\sqrt{b'-a'}$.
It is clear that $K$ annihilates the orthogonal complement of the space spanned by all such vectors. \hfill\SquareCastShadowTopRight\par\bigskip

%%=====================================================================================================================================================================

%=====================================================================================================================================================================
%lemma
%=====================================================================================================================================================================
\begin{lem}\label{lem 3.3} Let $a=-b$,\footnote{This assumption is made purely to simplify the formulae.} and take $f$ to be the characteristic function $\chi_{\left(a,b\right)}$ of $\left(a,b\right)$. The set of $b\in\left(0,\infty\right)$ for which $\lim_{z\rightarrow+\infty}\sin\theta_{f,z}=0$ is of Lebesgue measure zero.
\end{lem}
%=====================================================================================================================================================================
\noindent\textbf{\textit{Proof}:} Explicitly, we have
\begin{flalign*}
\mu_z:=|\sin\theta_{f,z}| = 
\sqrt{\frac{1}{2}\left(1-\frac{\hat{f}\left(0\right)m}{\lambda_z\Xi_{f,z}}\right)}.
\end{flalign*}
Using the formula \eqref{eq:Xifz} for $\Xi_{f,z}$, we easily see that $\mu_z\to 0$ only if $\lambda_z^2\,\lvert\hat{f}\left(2\lambda_z\right)\rvert^2\left(1-\frac{m^2}{\lambda_z^2}\right)\to 0$, which implies 
$\lambda_z^2\,\lvert\hat{f}\left(2\lambda_z\right)\rvert^2\to 0$, since $\lim_{z\rightarrow\infty}\lvert\lambda_z\rvert=\infty$. Now, for $f=\chi_{\left(-b,b\right)}$, we have $\hat{f}\left(2\lambda_z\right)=\sin\left(2b\lambda_z\right)/\lambda_z$ and so   $\lambda_z^2\,\lvert\hat{f}\left(2\lambda_z\right)\rvert^2=\sin\left(2b\lambda_z\right)^2$. It is proven in \cite{FeV12b} (after Proposition 4.1) that the set $\left\{b\in\left(0,\infty\right)\mid\lim_{z\rightarrow\infty}\sin\left(2b\lambda_z\right)=0\right\}$ is of Lebesgue measure zero.\hfill\SquareCastShadowTopRight\par\bigskip
%=====================================================================================================================================================================

%=====================================================================================================================================================================
%theorem
%=====================================================================================================================================================================
\begin{thm} Let \emph{$\textbf{M}$} be with spin connections as in Subsection \ref{subsec spatial Dirac operators} and with $a=-b$ for $b\in\left(0,\infty\right)$; hence, \emph{$\textbf{M}$} is a parallelisable, oriented and globally hyperbolic ultrastatic slab with compact spatial section. 
Then the unsoftened FP-state on the $C$\!*-completion of the self-dual CAR-algebra for the quantised free Dirac field fails to be Hadamard for
all $b\in(0,\infty)$ outside a set of Lebesgue measure zero. 
\end{thm}
%=====================================================================================================================================================================
%proof
%=====================================================================================================================================================================
\noindent\textbf{\emph{Proof:}} If $b\in\left(0,\infty\right)$ is such that $\lim_{z\rightarrow+\infty}\mu_z\neq0$, then $K$ cannot
be compact \cite[Thm.4.24(b)]{Ru91}. We conclude that the FP-state cannot be a Hadamard state for such choices of $b$,
and  Lemma \ref{lem 3.3} completes the proof. \hfill\SquareCastShadowTopRight\par\bigskip
%=====================================================================================================================================================================

%=====================================================================================================================================================================
%text
%=====================================================================================================================================================================
Note that the softened FP-states, for which $f$ is smooth and compactly supported, avoid the contradiction in the proof of the theorem due to the Riemann-Lebesgue lemma applied to (derivatives of) $f$. We will now show that such states are always Hadamard.
%=====================================================================================================================================================================

%=====================================================================================================================================================================
%subsection
%=====================================================================================================================================================================
\subsection{Softened FP-states are Hadamard}\label{subsec Brum-Fredenhagen}
%=====================================================================================================================================================================

The aim of this section is to prove the following result. 
%=====================================================================================================================================================================
%theorem
%=====================================================================================================================================================================
\begin{thm}\label{thm Brum-Fredenhagen is Hadamard} Let \emph{$\textbf{M}$} be with spin connections as in Subsection \ref{subsec spatial Dirac operators}; in particular, \emph{$\textbf{M}$} is a parallelisable, oriented and globally hyperbolic ultrastatic slab with compact spatial section. The softened FP-states on the $C$\!*-completion of the self-dual CAR-algebra for the quantised free Dirac field on \emph{$\textbf{M}$} are Hadamard.
\end{thm}
%=====================================================================================================================================================================
The proof is accomplished by the following discussion. Let $f$ be a smooth compactly supported and nonnegative function, not identically zero,
so the state $\omega_{\text{FP}_f}$ is a softened FP-state. To establish that the difference $W^{\left(2\right)}_{\text{FP}_f}-W^{\left(2\right)}_0$ is smooth, we can equivalently show that $u'\otimes v\longmapsto\langle S v^\dagger\mid\left(Q^\sp_f-Q^\sp\right)S u'\rangle_\sp$ for $u'\in\mathcal{C}^\infty_0\left(M,\mathbb{C}^4\right)$ and $v\in\mathcal{C}^\infty_0\left(M,\left(\mathbb{C}^4\right)^*\right)$ is smooth. We know that we can write 
\begin{flalign*}
&&\langle S v^\dagger\mid\left(Q^\sp_f-Q^\sp\right)S u'\rangle_\sp=\sum_{z\in\mathbb{Z}^+}\int\limits_{M\times M}\sigma_z^*\left(u'\otimes v\right)\vol_{\textbf{M}\times\textbf{M}}&&\\
&&&&\makebox[0pt][r]{$u'\in\mathcal{C}^\infty_0\left(M,\mathbb{C}^4\right)$, $v\in\mathcal{C}^\infty_0\left(M,\left(\mathbb{C}^4\right)^*\right)$,}
\end{flalign*}
where $\sigma_z\in\mathcal{C}^\infty\left(M\times M,\mathbb{C}^4\otimes\left(\mathbb{C}^4\right)^*\right)$ is read off from (\ref{difference two-point distributions}):
\begin{align}\label{sigma+z}
\sigma_z &= \gamma^0\kappa_{f,z}^+\otimes 
(\kappa_{f,z}^+)^\dagger \gamma^0 -  \gamma^0\kappa_{z}^+\otimes 
(\kappa_{z}^+)^\dagger \gamma^0 \nonumber \\
&=  \sin^2 \theta_{f,z} \left(\gamma^0\kappa_{z}^-\otimes 
(\kappa_{z}^-)^\dagger \gamma^0 - \gamma^0\kappa_{z}^+\otimes 
(\kappa_{z}^+)^\dagger \gamma^0\right) \nonumber \\
&\qquad + 
\sin  \theta_{f,z}\cos \theta_{f,z}
\left( e^{-\iu \phi_{f,z}}\gamma^0\kappa_{z}^-\otimes 
(\kappa_{z}^+)^\dagger \gamma^0 + e^{\iu \phi_{f,z}}\gamma^0\kappa_{z}^+\otimes 
(\kappa_{z}^-)^\dagger \gamma^0
\right) 
\end{align}

To show that $\omega_{\text{FP}_f}$ is Hadamard, it is sufficient to show that the series $\sum_{z\in\mathbb{Z}^+}\sigma_z$ converges in $L^2\left(M\times M,\mathbb{C}^4\otimes\left(\mathbb{C}^4\right)^*;\vol_{\textbf{M}\times\textbf{M}}\right)$ and has a smooth representative. 
Now $\sigma_z$ decomposes into four pairwise orthogonal terms, because the vectors $\chi_{sz}\otimes\chi_{s'z}$ ($s,s'\in\{\pm 1\}$) are
orthonormal in $L^2(\Sigma;\mathbb{C}^4;\vol_h)$ and so we compute
\begin{flalign}\label{eq:sigmazsq}
&&\lVert\sigma_z\rVert_2^2
=(b-a)\left(2\sin^4\theta_{f,z} + 2\sin^2\theta_{f,z}  \cos^2\theta_{f,z}\right)  = 2(b-a) \sin^2\theta_{f,z}   &&\makebox[0pt][r]{$\forall z\in\mathbb{Z}^+$.}
\end{flalign}
As the $\sigma_z$ are pairwise orthogonal for $z\in\mathbb{Z}^+$, their
sum converges if $\sum_{z\in\mathbb{Z}^+}\lVert\sigma_z\rVert^2_2<\infty$ (cf. \cite[Lem.21.7]{HS96}), which is established by the $p=0$ case of the following lemma.
%=====================================================================================================================================================================
%lemma
%=====================================================================================================================================================================
\begin{lem}\label{lem lemma} For each $p=0,1,2,\ldots$, 
\begin{equation}\label{sum}
\sum_{z\in\mathbb{Z}^+} \lambda_z^p \sin^2\theta_{f,z} <\infty.
\end{equation}  
\end{lem}
%=====================================================================================================================================================================
\noindent\textbf{\textit{Proof}:} Note first that all terms in the series are positive. It follows from Eq.~\eqref{eq:param} that
\begin{flalign*}
2  \sin^2\theta_{f,z} =\left(1-\frac{\hat{f}\left(0\right)m}{\lambda_z\Xi_{f,z}}\right).
\end{flalign*}
Using the explicit form of $\Xi_{f,z}$ found in Eq.~\eqref{eq:Xifz}, we compute
\begin{align*}
\sum_{z\in\mathbb{Z}^+}2\lambda_z^p \sin^2\theta_{f,z} 
&	=	\sum_{z\in\mathbb{Z}^+}\left( 1-\frac{1}{\sqrt{\lvert\hat{g}\left(2\lambda_z\right)\rvert^2\left(\lambda_z^2-m^2\right)+1}}\right)\lambda_z^p\displaybreak[0]\\
&	\leq	\sum_{z\in\mathbb{Z}^+}\lvert\hat{g}\left(2\lambda_z\right)\rvert^2 \lambda_z^{p+2}
\end{align*}
where $g:=f/\left(\hat{f}\left(0\right)m\right)\in\mathcal{C}^\infty_0\left(\mathbb{R},\mathbb{R}\right)$; also, recall that $\lambda_z^2-m^2\geq0$. 
Standard estimates (e.g., \cite[(8.1.1)]{Hoer1}) yield constants $C_N>0$ for $N\in\mathbb{N}$ such that $\lvert\hat{g}\left( \lambda \right)\rvert\leq C_N \left(1+ \lvert\lambda\rvert\right)^{-N}$ for all $\lambda\in\mathbb{R}$. Hence,
\begin{flalign*}
&&\sum_{z\in\mathbb{Z}^+} 2\lambda_z^p \sin^2\theta_{f,z}
	\leq	\sum_{z\in\mathbb{Z}^+}\frac{C^2_N}{\left(1+2 \lambda_z \right)^{2N}} \lambda_z^{p+2}
	\leq	\sum_{z\in\mathbb{Z}^+}\frac{C^2_N}{2^{2N}} \lambda_z^{p+2-2N},&&N\in\mathbb{N}.
\end{flalign*}
According to \cite[Chap.III, (5.6)]{LM89}, we know that there exists a constant $c>0$ such that $d\left(\Lambda\right)\leq c\Lambda^{21/2}$ holds for all $\Lambda>0$, where $d\left(\Lambda\right)=\dim\left(\bigoplus_{\lvert\lambda\rvert\leq\Lambda}E_\lambda\right)$ and $E_\lambda$ is the eigenspace of the smooth $\mathbb{C}^4$- resp. $\left(\mathbb{C}^4\right)^*$-valued eigenfunctions on $\Sigma$ to the eigenvalue $\lambda$ of $H^\sp$ (resp. $H^\cosp$). Let $M\left(z\right):=\max\left\{w\in\mathbb{Z}'\mid\lvert\lambda_w\rvert=\lvert\lambda_z\rvert\right\}$, which exists by the finite multiplicity of the eigenvalues of $H^\sp$ and $H^\cosp$; then by counting, we readily see $\lvert z\rvert\leq M\left(z\right)=d\left(\lvert\lambda_z\rvert\right)\leq c\,\lvert\lambda_z\rvert^{21/2}$ owing to the way we have ordered the countably many eigenvalues of $H^\sp$ and $H^\cosp$ (cf. the end of Subsection \ref{subsec spatial Dirac operators}). It follows that $\lvert\lambda_z\rvert\geq k|z|^{2/21}$, where $k=c^{-2/21}$. Letting $N>\frac{p}{2}+1$,
\begin{align*}
\sum_{z\in\mathbb{Z}^+}  2\lambda_z^p \sin^2\theta_{f,z}  
	\leq	\sum_{z\in\mathbb{Z}^+}\frac{C^2_N}{2^{2N} \lambda_z^{2N-p-2}}
	\leq	\sum_{z\in\mathbb{Z}^+}\frac{C^2_N}{2^{2N}k^{2N-p-2}{z^2}^{\left(2N-p-2\right)/21}},
\end{align*}
which converges for $N>p/2+25/4$. \hfill\SquareCastShadowTopRight\par\bigskip
%===================================================================================================================================================================== 

%===================================================================================================================================================================== 
%text
%===================================================================================================================================================================== 
With the $L^2$-convergence of $\sigma=\sum_{z\in\mathbb{Z}^+}\sigma_z$ now established, 
the remaining task is to show that $\sigma$ has a smooth representative. This will be accomplished by using Sobolev
estimates, for which it is convenient to embed $M$ in a $4$-dimensional compact Riemannian manifold $X$ defined
as follows. Reconsider the parallelisable, $3$-dimensional, oriented, connected and compact Riemannian manifold $\left(\Sigma,h,\left[\Omega\right]\right)$ with which $\textbf{M}$ and $\textbf{N}$ are constructed and the $\left[\Omega\right]$-oriented, $h$-orthonormal smooth global framing $\left(\eta_1,\eta_2,\eta_3\right)$. We also use $\tilde{\eta}_0$ to denote the unique smooth vector field on
the circle $S^1$ (regarded as the unit circle of $\mathbb{C}$)
satisfying $\tilde{\eta}_0\left(f\right)|_z=\frac{d}{dt}f\left(e^{it}z\right)|_{t=0}$ for all $z\in S^1$ and for all $f\in\mathcal{C}^\infty\left(S^1,\mathbb{R}\right)$. Then the standard Riemannian metric on $S^1$, $g_R$ may be defined by $g_R(\tilde{\eta}_0,\tilde{\eta}_0)=1$,
and the orientation $[\omega]$ so that $\omega(\tilde{\eta}_0)=1$. 
We define $X$ as the smooth product manifold $X =S^1\times\Sigma$, equipped with Riemannian metric 
$g_R:=\pr_1^*g_{S^1}+\pr_2^*h$. The triple $\left(X,g_R,\left[\pr_1^*\omega\wedge\pr_2^*\Omega\right]\right)$, 
constitutes $4$-dimensional, oriented, connected and compact Riemannian manifold.
 
Now consider the smooth embedding $j:\left(a,b\right)\longrightarrow S^1$, $t\longmapsto e^{2\pi\iu\left(t-a\right)/(b-a)}$
and the corresponding embedding $\psi:M\longrightarrow X$, $\psi=j\times\id_\Sigma$. 
Fix  $\chi\in \mathcal{C}^\infty_0\left(a,b\right)$ and define $\tau_z\in  \mathcal{C}^\infty(X\times X, \mathbb{C}^4\otimes\left(\mathbb{C}^4\right)^*)$ by the push-forward $\tau_z = (\psi\times\psi)_* (\chi\otimes\chi)\sigma_z$. 
Then it is easily seen that $(\chi\otimes\chi)\sigma=\sum_{z\in\mathbb{Z}^+} 
(\chi\otimes\chi)\sigma_z$ has a smooth representative in $L^2\left(M\times M,\mathbb{C}^4\otimes\left(\mathbb{C}^4\right)^*;\vol_{\textbf{M}\times\textbf{M}}\right)$ if and only if $\sum_{z\in\mathbb{Z}^+} 
\tau_z$ exists in  $L^2\left(X\times X,\mathbb{C}^4\otimes\left(\mathbb{C}^4\right)^*;\vol_{\textbf{M}\times\textbf{M}}\right)$
and has a smooth representative $\tau\in \mathcal{C}^\infty(X\times X, \mathbb{C}^4\otimes\left(\mathbb{C}^4\right)^*)$; 
indeed, the pull-back of $\tau$ by $\psi\times\psi$ is a smooth representative of $(\chi\otimes\chi)\sigma$. If
this can be done for arbitrary $\chi$, then $\sigma$ itself has a smooth representative -- this will now occupy the remainder 
of the section. 

We introduce two first order partial differential operators on $X$: 
\begin{flalign}\label{alternative elliptic linear differential operator I}
&&D^\text{s}u= \gamma^0\left(-\tilde{\eta}_0\otimes\bbone+\bbone\otimes H^\sp\right)\gamma^0u,&&u\in\mathcal{C}^\infty\left(X,\mathbb{C}^4\right), 
\end{flalign}
and
\begin{flalign}\label{alternative elliptic linear differential operator II}
&&D^\text{c}v=\left[\left(\tilde{\eta}_0\otimes\bbone+\bbone\otimes H^\cosp\right)v\gamma^0\right]\gamma^0,&& v\in\mathcal{C}^\infty\left(X,\left(\mathbb{C}^4\right)^*\right) , 
\end{flalign}
where $\bbone$ denotes the identity on $\mathcal{C}^\infty\left(S^1,\mathbb{C}\right)$ and we have made use of the standard continuous identifications.

%===================================================================================================================================================================== 

%===================================================================================================================================================================== 
%lemma
%===================================================================================================================================================================== 
\begin{lem} $D^\text{s}:\mathcal{C}^\infty\left(X,\mathbb{C}^4\right)\longrightarrow\mathcal{C}^\infty\left(X,\mathbb{C}^4\right)$ and $D^\text{c}:\mathcal{C}^\infty\left(X,\left(\mathbb{C}^4\right)^*\right)\longrightarrow\mathcal{C}^\infty\left(X,\left(\mathbb{C}^4\right)^*\right)$ are elliptic.
\end{lem}
%===================================================================================================================================================================== 
%proof
%===================================================================================================================================================================== 
\noindent\textbf{\textit{Proof}:} We show this claim for $D^\text{c}$; the proof for $D^\text{s}$ is analogous. From (\ref{alternative elliptic linear differential operator II}) we obtain for the principal symbol $\sigma_{D^\text{c}}\left(\xi\right)=\xi_0+\iu\xi_i\gamma^0\gamma^i$ (cf.\ Lemma~\ref{lem spatial Dirac operators are self-adjoint}); hence $\det\left(\xi_0+\iu\xi_i\gamma^0\gamma^i\right)=\left(\xi_0^2+\xi_1^2+\xi_2^2+\xi_3^2\right)^2$, which shows that $\sigma_{D^\text{c}}\left(\xi\right)$ is an isomorphism of complex vector spaces for all $\xi\in T^*\!X$ unless $\xi=0\in T^*\!X_{\left(t,x\right)}$ for $\left(t,x\right)\in X$.\hfill\SquareCastShadowTopRight\par\bigskip
%===================================================================================================================================================================== 

We will need to introduce various Sobolev spaces of vector-valued functions on $X$ and $X\times X$, 
each of which can be defined as the completion of the space of smooth vector-valued functions in an appropriate Sobolev norm. 
In fact, there are many equivalent norms that can be used: any linear connection determines a corresponding basic Sobolev norm
of order $s=0,1,2,\ldots$ \cite[Chap.III, \S2]{LM89} (and the norms induced by different connections and inner products are all
equivalent). However, any elliptic partial differential operator of order $s$ also induces an equivalent norm and therefore the
same completion \cite[Thm.III.5.2(iii)]{LM89}. We may therefore define the following Sobolev spaces, for $s\in \mathbb{N}_0$: 
$L^2_{s}(X,\mathbb{C}^4;\vol_X)$ and $L^2_{s}(X,(\mathbb{C}^4)^*;\vol_X)$ are defined to be the completions of $\mathcal{C}^\infty\left(X,\mathbb{C}^4\right)$ and $\mathcal{C}^\infty\left(X,(\mathbb{C}^4)^*\right)$ 
with respect to the norms defined by 
\begin{align*}
\left\lVert F \right\rVert_{X,s}^2 &:= \left\lVert F\right\rVert^2_{X,0} + \left\lVert\left(D^{\mathrm{s}}\right)^s F\right\rVert^2_{X,0} \\
\left\lVert G \right\rVert_{X,s}^2 &:= \left\lVert G\right\rVert^2_{X,0} + \left\lVert\left(D^{\mathrm{c}}\right)^s G\right\rVert^2_{X,0}
\end{align*}
for $F\in \mathcal{C}^\infty\left(X,\mathbb{C}^4\right)$, $G\in \mathcal{C}^\infty\left(X,(\mathbb{C}^4)^*\right)$
(we do not distinguish notationally between these norms, as it will always be clear which is intended), where $\left\lVert \cdot \right\rVert_{X,0}$ 
denotes the ordinary $L^2$-norm. Similarly, we define $L^2_{s}(X\times X,\mathbb{C}^4\otimes\left(\mathbb{C}^4\right)^*;\vol_{X\times X})$
to be the completion of 
$\mathcal{C}^\infty\left(X\times X,\mathbb{C}^4\otimes\left(\mathbb{C}^4\right)^*\right)$ with respect to the norm
\begin{align*}
\left\lVert H \right\rVert_{X\times X,s}^2 &:= \left\lVert H\right\rVert_{X\times X,0}^2 + 
\left\lVert\left( (D^{\mathrm{s}})^s\otimes \bbone+ \bbone\otimes (D^{\mathrm{c}})^s\right) H\right\rVert_{X\times X,0}^2  .
\end{align*}
Each of these spaces has a natural Hilbert space inner product compatible with the norms just given, e.g., 
\begin{equation*}
\ip{F}{F'}_{X,s} := \ip{F}{F'}_{X,0} + \ip{ \left(D^{\mathrm{s}}\right)^sF}{\left(D^{\mathrm{s}}\right)^sF'}_{X,0} 
\end{equation*}
for $F, F'\in\mathcal{C}^\infty\left(X,\mathbb{C}^4\right)$. The choices just made ensure that the various norms interact well: 
%=====================================================================================================================================================================
%lemma
%=====================================================================================================================================================================
\begin{lem}\label{lem Sobolev norm} 
For $s=0,1,2,\ldots$, we have the estimate
\begin{flalign*}
\lVert F\otimes G\rVert^2_{X\times X,s}\leq 2 \lVert F\rVert^2_{X,s}\lVert G\rVert^2_{X,s}
\end{flalign*}
for all $F\in \mathcal{C}^\infty\left(X,\mathbb{C}^4\right)$, $G\in \mathcal{C}^\infty\left(X,(\mathbb{C}^4)^*\right)$. 
\end{lem}
%=====================================================================================================================================================================
\noindent\textbf{\textit{Proof}:} We compute
\begin{align*}
\lVert F\otimes G\rVert^2_{X\times X,s} &=  \lVert F\rVert^2_{X,0}\lVert G\rVert^2_{X,0} + 
\lVert\left( \left(D^{\mathrm{s}}\right)^sF\right)\otimes G + F\otimes \left( \left(D^{\mathrm{c}}\right)^s G\right)\rVert^2_{X\times X,0} \\
&\le  \lVert F\rVert^2_{X,0}\lVert G\rVert^2_{X,0} + 
2\lVert\left( \left(D^{\mathrm{s}}\right)^sF\right)\otimes G \rVert^2_{X\times X,0}+ 2 \lVert F\otimes \left( \left(D^{\mathrm{c}}\right)^s G\right)\rVert^2_{X\times X,0} \\
&\le \lVert F\rVert^2_{X,0}\lVert G\rVert^2_{X,0} + 
2\lVert \left(D^{\mathrm{s}}\right)^sF\rVert^2_{X,0}  \lVert G \rVert^2_{X,0}+ 2 \lVert F\rVert^2_{X,0}  \lVert \left(D^{\mathrm{c}}\right)^s G\rVert^2_{X,0} \\
&\le 2 \lVert F\rVert^2_{X,s}\lVert G\rVert^2_{X,s},
\end{align*}
where we have used the Hilbert space inequality $\|a+b\|^2\le 2\|a\|^2+2\|b\|^2$ (a consequence of the parallelogram law). 
\hfill\SquareCastShadowTopRight
%=====================================================================================================================================================================

Next, note that, for $u\in \mathcal{C}^\infty(S^1,\mathbb{C})$ we have
\begin{flalign*}
&&D^{\mathrm{s}}\left[ u\gamma^0\chi_z\right]
=\left(-\tilde{\eta}_0\left(u\right)+u\lambda_z\right)\gamma^0\chi_z, && z\in\mathbb{Z}'
\end{flalign*}
and
\begin{flalign*}
&&D^\mathrm{c}\left[u\zeta_z\gamma^0\right]=\left(\tilde{\eta}_0\left(u\right)+u\lambda_z\right)\zeta_z\gamma^0, && z\in\mathbb{Z}'
\end{flalign*}
and we obtain by induction
\begin{flalign*}
&&
\left(D^{\mathrm{s}}\right)^{s}\left[u\gamma^0\chi_z\right]
=(-1)^s (P_{-\lambda_z,s}u)\gamma^0\chi_z, \qquad
\left(D^\mathrm{c}\right)^s\left[u\zeta_z\gamma^0\right]= (P_{\lambda_z,s}u) \zeta_z\gamma^0, && z\in\mathbb{Z}'
\end{flalign*}
where $P_{\lambda,s}$ ($\lambda\in\mathbb{R}$, $s\in\mathbb{N}_0$) 
is a differential operator on $\mathcal{C}^\infty(S^1)$ of order $s$, depending polynomially on $\lambda$, given by 
\begin{equation*}
P_{\lambda,s}u:= \sum_{k=0}^s\binom{s}{k} \tilde{\eta}^{s-k}_0\left(u\right)\lambda^k.  
\end{equation*}
A number of consequences follow. First, we see that
\begin{equation*}
\lVert u\gamma^0\chi_z \rVert^2_{X,s}  = \|u\|^2_{S^1,0} +  \|P_{-\lambda_z,s} u\|^2_{S^1,0},  
\qquad 
\lVert v\zeta_z \gamma^0\rVert^2_{X,s}= \|v\|^2_{S^1,0} +  \|P_{\lambda_z,s} v\|^2_{S^1,0}
\end{equation*}
and hence
\begin{equation*}
\lVert (u\gamma^0\chi_z)\otimes(v\zeta_w \gamma^0) \rVert^2_{X\times X,s}
\le 2\left( \|u\|^2_{S^1,0} +  \|P_{-\lambda_z,s} u\|^2_{S^1,0}\right)\left( \|v\|^2_{S^1,0} +  \|P_{\lambda_w,s} v\|^2_{S^1,0}\right)
\end{equation*}
for all $u,v\in \mathcal{C}^\infty(S^1,\mathbb{C})$, $w,z\in\mathbb{Z}'$. 

Let us also observe that, for $\lambda\in\mathbb{R}$, $s\in\mathbb{N}_0$, $Q_{\lambda,s}:=
|P_{\lambda,s} j_*(\chi e^{\iu \lambda \cdot})|^2$ is  
a polynomial of degree $2s$ in $\lambda\in\mathbb{R}$ with nonnegative coefficients in $\mathcal{C}^\infty(S^1,\mathbb{R})$,
whose $L^1$-norm $\int_{S^1} Q_{\lambda,s}\vol_{S^1}$ is also a polynomial of the same degree. 
Therefore there is, for each $s\in\mathbb{N}_0$, a constant $C_s>0$ such that
\begin{equation*}
\| j_*(\chi e^{ \iu \lambda \cdot})\|_{S^1,0}^2+ \| P_{\lambda,s} j_*(\chi e^{ \iu \lambda \cdot})\|_{S^1,0}^2
\le C_s (1+|\lambda|^{2s})
\end{equation*}
for all $\lambda\in\mathbb{R}$. From this, it follows that both 
$\| \psi_*\chi \gamma^0\kappa_z^\pm \|^2_{X,s}$ and 
$\| \psi_*\chi \kappa_z^{\pm\dagger}\gamma^0\|^2_{X,s}$ are bounded above by $C_s  (1+|\lambda_z|)^{2s}$ 
for all $z\in\mathbb{Z}^+$ and hence
\begin{equation*}
\| (\psi\times\psi)_*(\chi \gamma^0\kappa_z^\pm) \otimes (\chi \kappa_z^{\pm\dagger}\gamma^0)
\|^2_{X\times X,s}\le 2 C_s^2 (1+\lambda_z^{2s})^2 
 \end{equation*}
for arbitrary and independent choices of the signs on the left-hand side. 
The function $\tau_z$ is a linear combination of four such terms, which are mutually orthogonal in
$L^2_{s}(X\times X,\mathbb{C}^4\otimes\left(\mathbb{C}^4\right)^*;\vol_{X\times X})$.
Thus (cf.\ the computation \eqref{eq:sigmazsq}) 
\begin{flalign*}
&&\| \tau_z\|^2_{X\times X,s}\le 4 C_s^2 (1+\lambda_z^{2s})^2  \sin^2\theta_{f,z}, &&z\in\mathbb{Z}^+.
\end{flalign*}
Owing to Lemma \ref{lem lemma}, 
we may conclude, for each $s=0,1,2,\ldots$, that
$\sum_{z\in\mathbb{Z}^+}\lVert\tau_z\rVert_{X\times X,s}^2<\infty$, because the $\tau_z$ are easily seen to be pairwise orthogonal in 
$L^2_{s}(X\times X,\mathbb{C}^4\otimes\left(\mathbb{C}^4\right)^*;\vol_{X\times X})$.
Accordingly, $\tau:=\sum_{z\in\mathbb{Z}^+}\tau_z$ not only exists in $L^2(X\times X,\mathbb{C}^4\otimes\left(\mathbb{C}^4\right)^*;\vol_{X\times X})$, but also has a smooth representative as a result of the Sobolev embedding theorem \cite[Thm.III.5.2(iii)]{LM89}. 
This concludes the proof of Theorem~\ref{thm Brum-Fredenhagen is Hadamard}, as $\chi\in\mathcal{C}_0^\infty(a,b)$ was arbitrary in the discussion above.  

%======================================================================================================================================================================
\subsection{Quantum fluctuations in the FP-state}\label{sec quantum fluctuations}
%======================================================================================================================================================================

In this subsection use FP-states to define Wick polynomials by 
normal ordering in Hilbert space representations, and then study
the fluctuations of such Wick polynomials in the FP-state.  We let $\textbf{M}$ again be an oriented and globally hyperbolic ultrastatic slab of dimension $4$ with compact spatial section $\Sigma$ and spin connections $\nabla^\sp$ and $\nabla^\cosp$ as in Section \ref{subsec spatial Dirac operators}. Let $f\in L^1(\mathbb{R})$
be nonnegative and consider the GNS representation
of the corresponding FP-state $\omega_{\text{FP}_f}$. The GNS
Hilbert space is a Fock space generated by creation and annihilation operators $b^\dagger_z,d^\dagger_z,b_z$ and $d_z$ ($z\in\mathbb{Z}^+$) obeying the anticommutation relations
$\{b_w,b_z^\dagger\}=\{d_w,d_z^\dagger\}=\delta_{wz}\bbone$ and all other anticommutators among these operators vanishing, 
and the GNS vector is the Fock vacuum vector $\Omega$. The
quantum Dirac spinor field and its Dirac adjoint take the form
\begin{flalign}\label{eq:psi_Fock}
&&
\Psi\left[v\right]&=\sum_{z\in\mathbb{Z}^+} \int_M \left( v\left(t,x\right)\kappa^+_{f,z}\left(t,x\right) b_z  
+v\left(t,x\right)\kappa^-_{f,z}\left(t,x\right)  d_z^\dagger
\right)
\vol_\textbf{M},
&&\\\nonumber
&&&&\makebox[0pt][r]{$v\in\mathcal{C}_0^\infty\left(X,(\mathbb{C}^4)^*\right)$,}
\end{flalign}
and
\begin{flalign}
&&
\Psi^\dagger\left[u\right]&=\sum_{z\in\mathbb{Z}^+} \int_M \left( 
\kappa^{+\dagger}_{f,z}\left(t,x\right) u\left(t,x\right) b_z^\dagger  
+\kappa^{-\dagger}_{f,z}\left(t,x\right) u\left(t,x\right) d_z
\right)
\vol_\textbf{M},
&&\\\nonumber
&&&&\makebox[0pt][r]{$v\in\mathcal{C}_0^\infty\left(X,\mathbb{C}^4\right)$.}
\end{flalign}
As a consistency check, it
is easily verified that $\langle\Omega\mid\left(\Psi\left[v\right]\Psi^\dagger\left[u'\right]+
\Psi^\dagger\left[u\right]\Psi\left[v'\right]\right)\Omega\rangle=
W^{\left(2\right)}_{\text{FP}_f}\left[\left(u\oplus v\right)\otimes\left(u'\oplus v'\right)\right]$ for all $u,u'\in\mathcal{C}_0^\infty\left(X,\mathbb{C}^4\right)$ and for all $v,v'\in\mathcal{C}_0^\infty\left(X,(\mathbb{C}^4)^*\right)$.
For comparison, in a Fock space representation where the Fock vacuum $\Omega_0$ represents the reference state $\omega_0$, i.e., the ground state, the quantum Dirac spinor field and the quantum Dirac cospinor field are given by the above
expressions but with $\kappa^\pm_{f,z}$ replaced by $\kappa^\pm_z$ respectively,
and yield the ground state two-point function $W^{\left(2\right)}_0$ in the analogous fashion. (If one replaces $\kappa^\pm_{f,z}$ replaced by $\kappa^\mp_z$, one obtains a `ceiling state'.)
%======================================================================================================================================================================

With normal ordering defined in the usual way, we define the energy density operator
\begin{flalign*}
\varrho:=\frac{i}{2}\left( \boldsymbol{\colon}\Psi^\dagger\gamma^0\dot{\Psi}-\dot{\Psi}^\dagger\gamma^0\Psi\boldsymbol{\colon}\right),
\end{flalign*} 
which has vanishing expectation value in the FP-state vector $\Omega$, by the definition of normal ordering.
We compute the fluctuations of $\varrho$ in $\Omega$, starting with the observation that
\begin{flalign}\label{eq D40}
&&\varrho\left(h\otimes g\right)\Omega&=\frac{i}{2}\sum_{w,z\in\mathbb{Z}^+} 
\int\limits_M \left( \kappa^{+*}_{f,w}\left(t,x\right)\dot{\kappa}^{-}_{f,z}\left(t,x\right) - 
\dot{\kappa}^{+*}_{f,w}\left(t,x\right)\kappa^{-}_{f,z} \left(t,x\right)
\right) h\left(t\right)g\left(x\right)\,
\vol_\textbf{M}
b^\dagger_w d^\dagger_z\Omega &&
\nonumber\\
&&&&\makebox[0pt][r]{$h\in\mathcal{C}_0^\infty(\mathbb{R} ,\mathbb{C})$, $g\in\mathcal{C}^\infty\left(\Sigma,\mathbb{C}\right)$.}
\end{flalign}
Taking $g$ to be the constant unit function on $\Sigma$, we can perform the integration over $\Sigma$ and make use of the orthonormality relations of the $\chi_z$. Noting, for example, that
\begin{align*}
\frac{i}{2}\int\limits_M  \kappa^{+*}_{f,w}\left(t,x\right)\dot{\kappa}^{-}_{f,z}\left(t,x\right)h\left(t\right)\,
\vol_\textbf{M}&= -\frac{\lambda_z}{2}\int\limits_M   
\left(\cos\theta_{f,w}\kappa^{+*}_w\left(t,x\right) + e^{i\phi_{f,w}}\sin\theta_{f,w}\kappa^{-*}_w\left(t,x\right)\right)\\
&\qquad\times
\left(\cos\theta_{f,z}\kappa^{-}_z\left(t,x\right) + e^{i\phi_{f,z}}\sin\theta_{f,z}\kappa^{+}_z\left(t,x\right)\right)\vol_\textbf{M} \\
&= -\frac{\lambda_z}{2} e^{i\phi_{f,z}} \cos\theta_{f,w}\sin\theta_{f,z}\delta_{wz}
 -\frac{\lambda_z}{2} e^{i\phi_{f,w}} \cos\theta_{f,z}\sin\theta_{f,w}\delta_{wz} \\ &= -\lambda_w e^{i\phi_{f,w}} \cos\theta_{f,w}\sin\theta_{f,w}\delta_{wz} ,
\end{align*}
we obtain for \eqref{eq D40}:
\begin{flalign*} 
&&\varrho\left(h\otimes 1\right)\Omega=-\hat{h}(0)\sum_{w\in\mathbb{Z}^+} \lambda_w e^{-i\phi_{f,w}}\sin 2\theta_{f,w}  b^\dagger_w d^\dagger_w\Omega &&
\makebox[0pt][r]{$h\in\mathcal{C}_0^\infty(\mathbb{R},\mathbb{C})$.}
\end{flalign*}
The operator $\varrho\left(h\otimes 1\right)$ is of course the total energy, up to an overall constant; the fact that the right-hand side depends on $h$ only via $\hat{h}(0)$
shows that the energy is conserved (note that $\Omega$ is not, in general, invariant under the time evolution).
The squared fluctuation of $\varrho\left(h\otimes 1\right)$ in the FP-state $\omega_{\text{FP}_f}$ is 
\begin{flalign}\label{eq D41}
\|\varrho\left(h\otimes 1\right)\Omega\|^2 = |\hat{h}(0)|^2 \sum_{w\in\mathbb{Z}^+} \lambda_w^2  \sin^2 2\theta_{f,w}
\end{flalign} 
(recall that $\rho(h\otimes 1)$ has vanishing expectation value in state $\Omega$).
Evidently the right-hand side of \eqref{eq D41} can only converge if $\sin^2 2\theta_{f,w}\to 0$ as $w\to\infty$ (faster than $\lambda_w^{-2}$)
and, hence, only if $\sin \theta_{f,w}\to 0$. In the case of the unsoftened FP-state, Lemma \ref{lem 3.3} yields:
%======================================================================================================================================================================
\begin{thm} Let $\textbf{M}$ be an oriented and globally hyperbolic ultrastatic slab of dimension $4$ with compact spatial section $\Sigma$, $a=-b$ for $b\in\left(0,\infty\right)$ and spin connections $\nabla^\sp$ and $\nabla^{\cosp}$ as in Section \ref{subsec spatial Dirac operators}. Then the set of the $b\in\left(0,\infty\right)$ for which the normal ordered energy density $\varrho$ has finite quantum fluctuations in the corresponding (i.e., unsoftened) FP-state is of Lebesgue measure zero.
\end{thm}
%======================================================================================================================================================================
We make no statement for the case where $b\in\left(0,\infty\right)$ is taken from the set of Lebesgue measure zero. In the case of the softened FP-states,
however, the fluctuations are finite, due to the estimate $\sin^2 2\theta\le 4 \sin^2\theta$, combined with the $p=2$ case of Lemma~\ref{lem lemma}. 

The calculation above is easily extended to deal with other Wick products. For example, it is easily seen that if $\varrho$ is replaced by 
\begin{flalign*}
\varrho^{(p)}:=\frac{i}{2}\left( \boldsymbol{\colon}\Psi^\dagger\gamma^0 \partial_t^{2p-1}\Psi-(\partial_t^{2p-1}\Psi)^\dagger\gamma^0\Psi\boldsymbol{\colon}\right),
\end{flalign*} 
for $p\in\mathbb{Z}^+$,
then the squared fluctuation of $\varrho^{(p)}\left(h\otimes 1\right)$ in the 
state $\omega_{\text{FP}_f}$ is  
\begin{flalign}\label{eq D42}
\|\varrho^{(p)}\left(h\otimes 1\right)\Omega\|^2 = |\hat{h}(0)|^2 \sum_{w\in\mathbb{Z}^+} \lambda_w^{4p-2}  \sin^2 2\theta_{f,w}.
\end{flalign} 
If $f$ is smooth and compactly supported in $(a,b)$, then the series converges for all $p\in\mathbb{Z}^+$
by Lemma~\ref{lem lemma}, and the fluctuations are finite. This is to be expected, in the light of Theorem~\ref{thm Brum-Fredenhagen is Hadamard}:
any product of smeared Wick polynomials defined relative to Hadamard states has a finite expectation
value in a Hadamard state -- a result going back to  
of Brunetti, Fredenhagen and K{\"o}hler~\cite{BruFreKoe96} in the scalar case, and (modulo inessential differences)
to Dappiaggi, Hack and Pinamonti~\cite{DapHacPin09} in the Dirac case.  However, we may also reverse the
reasoning: suppose all Wick polynomials  
have finite fluctuations in $\omega_{\text{FP}_f}$,
for some nonnegative $f\in L^1(\mathbb{R})$. 
Then the series on the right-hand side of \eqref{eq D42} converge for all $p\in\mathbb{Z}^+$.
By the remark above Lemma~\ref{lem spectrum of A}, we have $\cos^2\theta_{f,z}>1/2$ and hence $\sin^2 2\theta_{f,z}\ge 2\sin^2\theta_{f,z}$. 
Consequently  $\sum_{w\in\mathbb{Z}^+} \lambda_w^{4p-2}  \sin^2 \theta_{f,w}$ converges for each $p\in\mathbb{Z}^+$ and so the statement of Lemma~\ref{lem lemma} is valid for our choice of $f$ (whether or not it is actually smooth and compactly supported). Inspecting the proof of Theorem~\ref{thm Brum-Fredenhagen is Hadamard}, we see that it requires no other property of $f$, and so its statement also holds. Hence we have proved:
\begin{thm} \label{thm:fluc}
Let $\textbf{M}$ be an oriented and globally hyperbolic ultrastatic slab of dimension $4$ with compact spatial section $\Sigma$, $a,b\in\mathbb{R}$ with $a<b$, and spin connections $\nabla^\sp$ and $\nabla^{\cosp}$ as in Section \ref{subsec spatial Dirac operators}. 
If $f$ is any integrable nonnegative function such that the corresponding FP-state $\omega_{\text{FP}_f}$ has finite fluctuations for all its Wick polynomials,
then $\omega_{\text{FP}_f}$ is Hadamard. 
\end{thm}
The above result is an analogue of a result proved as \cite[Thm 2.3]{FeV13} for the scalar field, which asserts that general pure quasifree states with finite fluctuations for their associated Wick polynomials are
Hadamard. Here, we have proved the same statement for FP-states.
More generally, we conjecture that any pure quasifree state of the Dirac field on an ultrastatic slab that has
finite fluctuations for its Wick polynomials is either Hadamard or \emph{anti-Hadamard}. Here, a slight digression is needed: Recall that a Hadamard state may be defined in terms of the wave-front set \cite{Hoer1} of its two-point function, which gives
an equivalent formulation to the original definition using series expansions~\cite{KW91}.
Using a nonstandard Fourier convention (in line with that used elsewhere in this paper and in~\cite{R96}) the Hadamard condition may be simplified
to the requirement that $\text{WF}(W^{(2)})\subset \mathcal{N}^+\times\mathcal{N}^-$,
where $\mathcal{N}^{+/-}$ is the bundle of future/past-directed null covectors
on the spacetime $\textbf{M}$.\footnote{For equivalence of this condition with the original definition~\cite{KW91} in the scalar case, see \cite{R96} and~\cite{SV01} for a treatment encompassing fermionic fields as well.} 
By contrast, we say that the state is anti-Hadamard if $\text{WF}(W^{(2)})\subset \mathcal{N}^-\times\mathcal{N}^+$.
Thus, while Hadamard two-point functions are positive frequency in the
first slot and negative frequency in the second, the situation is 
reversed for anti-Hadamard states. 

The scalar field does not admit anti-Hadamard states, because the wave-front set condition turns out to be incompatible with the positivity of states, and this probably accounts for the lack of discussion of such states in the literature. However, anti-Hadamard states do exist for the  Dirac field -- the simplest
example is
the ceiling state\footnote{See \cite[Def.~5.3.18, Example~5.3.20]{BR_ii:97} for general results concerning ceiling states on CAR algebras.} on the full ultrastatic spacetime, which corresponds to 
replacing the mode functions $\kappa^\pm_{f,z}$ by $\kappa_z^\mp$ in \eqref{eq:psi_Fock}.
Anti-Hadamard states are excluded from the conclusion of
Theorem~\ref{thm:fluc} due to our assumption
that $f$ is nonnegative; reversing this choice, the resulting softened FP-states would be anti-Hadamard by the analogue
of Theorem~\ref{thm Brum-Fredenhagen is Hadamard}, using the ceiling state as the reference anti-Hadamard state.

While our conjecture for Dirac fields does not single out the Hadamard class as cleanly as is the case for the scalar field, it would still show that the finite fluctuation condition is linked to the ultraviolet behaviour of the two-point function, expressed via the wave-front set. 
Demanding additionally that the renormalised energy density obeys quantum energy inequalities (QEIs) \cite{FeV02} would then
select the Hadamard class, because anti-Hadamard states obey reversed QEIs in which averaged energy densities
are bounded from above. 

%=====================================================================================================================================================================

%=====================================================================================================================================================================
\section{Summary}\label{sec summary}
%=====================================================================================================================================================================
We have shown how to obtain a gauge invariant, pure and quasifree state, the FP-state, on the $C$*-completion of the self-dual CAR-algebra for the quantised free Dirac field following the covariant description in \cite{FR14a} of the fermionic projector. In our calculations, we have restricted ourselves to oriented globally hyperbolic ultrastatic slabs with compact spatial section and parameters $m>0$. We have shown that the FP-state suffers from the same shortcomings as the SJ-state for the quantised free real scalar field, that is, it can almost always be ruled out that the FP-state is a Hadamard state. 
Our arguments here are remarkably similar in spirit to those used in the case of the SJ-state \cite{FeV12b}. In view of this, the FP-state is as `badly' behaved (regarding the Hadamard property) as the SJ-state.

However, as we have also shown, the fermionic projector description is still valuable since a Hadamard state can always be obtained by a modification in the style of \cite{BF14}. In this way, the covariant character of the fermionic projector description is spoiled since a smooth cut-off function is introduced. It could perhaps be said that the FP-state is better behaved than the SJ-state because 
the smooth cut-off function appears only once in our construction, whereas it appears twice in the the Brum-Fredenhagen modification of the SJ-state (compare \cite[(19)]{BF14} with our \eqref{eq:Af}). Like \cite{BF14}, we have not investigated the detailed physical interpretation of the modified FP-state and leave this to a further analysis. However, they are of interest, at least as a class of Hadamard states constructed without explicit reference to the ultrastatic ground state or using the technique of spacetime deformation. Moreover, within the class of FP-states, there is the same tight link
between finiteness of fluctuations for Wick polynomials and the Hadamard condition as obtains for pure quasifree states of the scalar field
established in~\cite{FeV13}.

It would also be very interesting to see how our calculations and results can be carried over to globally hyperbolic spacetimes of finite lifetime that are not ultrastatic slabs or have non-compact spatial section. Similarly, reference \cite{FR14b} discusses the fermionic projector for spacetimes
of infinite lifetime and it would be interesting to extract states from that description and investigate their properties, though this would 
involve a much more complicated analysis.  
%=====================================================================================================================================================================

%=====================================================================================================================================================================
\medskip
\noindent \emph{Acknowledgement}
%=====================================================================================================================================================================
B.L. gratefully acknowledges the financial support of the Department of Mathematics, University of York, by the award of a \emph{Teaching Studentship}. C.J.F. thanks Bernard Kay, Rainer Verch and Felix Finster for useful discussions and correspondence.
%=====================================================================================================================================================================

%=====================================================================================================================================================================
%bibliography
%=====================================================================================================================================================================
\printbibliography
%=====================================================================================================================================================================

%======================================================================================================================================================================
\end{document}